\pdfoutput=1
\documentclass[acmsmall,screen,nonacm]{acmart}

\AtBeginDocument{%
  }

\setcopyright{acmlicensed}
\copyrightyear{2018}
\acmYear{2018}
\acmDOI{XXXXXXX.XXXXXXX}
\acmConference[Conference acronym 'XX]{Make sure to enter the correct
  conference title from your rights confirmation email}{June 03--05,
  2018}{Woodstock, NY}
\acmISBN{978-1-4503-XXXX-X/2018/06}

\usepackage{booktabs}   %
\usepackage{subcaption} %

\usepackage{marvosym}

\input{setup}

\begin{document}

\title[]{High-Level Synthesis of Efficient Pipelines with Visibility Control}

\author{Jungin Rhee}
\affiliation{%
	\institution{KAIST}
	\city{Daejeon}
	\country{Korea}
}
\email{jungin.rhee@kaist.ac.kr}

\author{Minseong Jang}
\affiliation{%
	\institution{KAIST}
	\city{Daejeon}
	\country{Korea}
}
\email{minseong.jang@kaist.ac.kr}

\author{Jaewoo Kim}
\affiliation{%
	\institution{Seoul National University}
	\city{Seoul}
	\country{Korea}
}
\email{jaewoo.kim.2@sf.snu.ac.kr}

\author{Jeehoon Kang}
\orcid{0000-0002-2115-0871}
\affiliation{%
	\institution{FuriosaAI}
	\city{Seoul}
	\country{Korea}
}
\email{jeehoon.kang@furiosa.ai}

\renewcommand{\shortauthors}{Rhee et al.}

\begin{abstract}
	{High-level synthesis (HLS) raises the abstraction of hardware design from concurrent register-transfer level (RTL).
Among the forms of parallelism HLS exploits, \emph{pipelining} demands fine-grained control over pipeline structure and hazard resolution to achieve competitive power, performance, and area (PPA).
However, existing tools either lack such control or sacrifice sequential semantics to provide it.

We present an HLS tool that embeds fine-grained pipeline control in a sequential programming model while generating synchronization logic only where needed to resolve the pipeline's potential hazards.
The tool builds on \emph{visibility control}, a novel programming abstraction that unifies hazard resolution techniques including stalling, bypassing, speculation, deferred commit, and register renaming.
We evaluate on in-order RISC-V cores, histograms, and an AES accelerator.
On RISC-V cores, we implement stall, bypass, speculation, and register renaming; on histograms, we implement scheduling strategies that previously required RTL or concurrent programming models.
Compiled pipelines outperform HLS tools with sequential semantics and achieve PPA comparable to hand-written RTL.
}
\end{abstract}


\maketitle

{\section{Introduction}
\label{sec:intro}

High-level synthesis (HLS) improves hardware design productivity by compiling imperative specifications into register-transfer-level (RTL) implementations.
A central challenge in HLS is to exploit hardware parallelism for performance while offering a simple programming model.

Pipelining is a fundamental technique in hardware design for exploiting instruction-level parallelism (ILP) by overlapping the execution of successive operations.
This requires two design decisions that together define the \emph{pipelining strategy}:
\begin{enumerate*}
	\item the \emph{pipeline structure} specifies the number of stages and their boundaries, and
	\item the \emph{hazard resolution strategy} specifies how to synchronize accesses to shared state across overlapping operations by combining techniques such as stalling, bypassing, speculation, deferred commit, and register renaming.
\end{enumerate*}

The pipelining strategy is a major determinant of a design's power, performance, and area (PPA), making it a primary target of design-space exploration.
Adding pipeline stages shortens the critical path and permits a higher clock frequency, but the additional pipeline registers increase area and power.
Likewise, more aggressive hazard resolution reduces stalls and improves throughput, but it increases area and power and may lower the achievable clock frequency.

Existing HLS approaches struggle to let designers easily write pipelines whose strategies suit the target workload and PPA budget.
Specifically, they have limitations in one or more of the following respects~(\cref{intro:tab:comparison}):
\begin{enumerate*}
	\item \emph{sequential source semantics}, which spare designers from managing concurrency,
	\item \emph{fine-grained control} over both pipeline structure and hazard resolution strategy, and
	\item \emph{pipeline-optimized synchronization logic}, in which the compiler analyzes the pipeline's potential hazards and generates synchronization logic only where needed to resolve them.
\end{enumerate*}

\begin{table}[t]
	\caption{Pipelining approaches in HLS (a) and hazard resolution expressiveness (b).}

	\label{intro:tab:comparison}
	\centering
	\footnotesize

	\begin{subtable}[t]{\linewidth}
		\caption{
			Pipelining approaches in HLS.
			\pmark{}: Partial support.
			\cmark$^{a}$: Subject to path-sensitive code rules (\cref{background:pdl-limitations}).
		}
		\label{intro:tab:hls_comparison}
		\setlength{\tabcolsep}{2.5pt}
		\renewcommand{\arraystretch}{1.1}
		\renewcommand{\tabularxcolumn}[1]{m{#1}}
		\begin{tabularx}{\linewidth}{|>{\raggedright\arraybackslash}m{0.25\linewidth}||>{\centering\arraybackslash}m{0.08\linewidth}|*{2}{>{\centering\arraybackslash}X|}>{\centering\arraybackslash}m{0.25\linewidth}|}
			\hline
			\multirow[c]{2}{=}{\centering\textbf{Pipelining approach}} & \multirow[c]{2}{=}{\makecell[c]{\textbf{Seq.}                                                     \\\textbf{source}}} & \multicolumn{2}{c|}{\textbf{Pipelining strategy control}} & \multirow[c]{2}{=}{\centering\textbf{Pipeline-optimized synchronization logic}} \\
			\cline{3-4}
			                                                           &                                               & \textbf{Structure} & \textbf{Hazard resolution} & \\
			\hline\hline
			Automated static scheduling~\cite{vitis,catapult}
			                                                           & \makecell[c]{\cmark}
			                                                           & \makecell[c]{\pmark                                                                               \\[-2pt]\scriptsize (by constraints)}
			                                                           & \xmark
			                                                           & \makecell[c]{\textemdash                                                                          \\[-2pt]\scriptsize (no runtime synchronization)} \\
			\hline
			Automated dynamic scheduling~\cite{dynamatic,spechls}
			                                                           & \makecell[c]{\cmark}
			                                                           & \xmark
			                                                           & \xmark
			                                                           & \makecell[c]{\cmark}                                                                              \\
			\hline
			Wavefront Threading\newline (Kanagawa~\cite{kanagawa})
			                                                           & \makecell[c]{\xmark}
			                                                           & \makecell[c]{\pmark                                                                               \\[-2pt]\scriptsize (by constraints)}
			                                                           & \makecell[c]{\cmark}
			                                                           & \makecell[c]{\xmark                                                                               \\[-2pt]\scriptsize (user-written)} \\
			\hline
			Hazard lock + speculation\newline + exception API (PDL~\cite{pdl,xpdl})
			                                                           & \makecell[c]{\cmark$^{a}$}
			                                                           & \makecell[c]{\cmark}
			                                                           & \makecell[c]{\pmark                                                                               \\[-2pt]\scriptsize (\cref{intro:tab:granularity})}
			                                                           & \makecell[c]{\xmark                                                                               \\[-2pt]\scriptsize (blackbox module per API)} \\
			\hline
			\hline
			\textbf{Visibility control (Ours)}
			                                                           & \makecell[c]{\cmark}
			                                                           & \makecell[c]{\cmark}
			                                                           & \makecell[c]{\cmark}
			                                                           & \makecell[c]{\cmark}                                                                              \\
			\hline
		\end{tabularx}
	\end{subtable}

	\vspace{2mm}
	\begin{subtable}[t]{\linewidth}
		\caption{hazard resolution expressiveness, along four dimensions.
			\cmark$^{a}$: Policy selectable on each access.
		}
		\label{intro:tab:granularity}
		\centering
		\begin{tabular}{|l||c|c|c|c|}
			\hline
			\textbf{Mechanism}   & \textbf{Per-state} & \textbf{Resolution policy} & \textbf{Per-address} & \textbf{Selectable obs. stage} \\
			\hline\hline
			Kanagawa shared var. & \cmark             & \cmark$^a$                 & \cmark               & \cmark                         \\
			\hline
			PDL hazard lock      & \cmark             & Conservative only          & \cmark               & \cmark                         \\
			\hline
			PDL speculation      & \xmark             & Speculative only           & \xmark               & \xmark                         \\
			\hline
			\hline
			\textbf{Var (Ours)}  & \cmark             & \cmark$^a$                 & \cmark               & \cmark                         \\
			\hline
		\end{tabular}
	\end{subtable}
\end{table}

Automated static scheduling, as used by Vitis HLS~\cite{vitis} and Catapult~\cite{catapult}, pipelines sequential code with a conservative schedule fixed at compile time that avoids hazards in all possible executions.
The designer can specify only general scheduling constraints, such as the initiation interval,
while the compiler determines the pipeline structure and instruction schedule.
While this approach works well for regular, statically analyzable dependencies,
it can severely limit throughput for designs with data-dependent control flow and memory accesses.

Automated dynamic scheduling, as used by Dynamatic~\cite{dynamatic} and SpecHLS~\cite{spechls}, provides runtime hazard resolution but gives designers no direct control over the pipeline structure or resolution strategy.
The compiler selects the pipeline stage boundaries and the hazard resolution mechanisms, and generates the corresponding stall, bypass, and speculation logic.
This lack of control prevents designers from exploring alternative pipelining strategies, resulting in substantially worse PPA than hand-written RTL pipelines.

Kanagawa~\cite{kanagawa} adopts a multithreaded programming model, termed Wavefront Threading, under which overlapping accesses to shared state need not behave as if executed sequentially.
Designers must therefore explicitly write synchronization logic for hazard resolution in the source.
This model gives designers fine-grained control over hazard resolution along four dimensions~(\cref{intro:tab:granularity}):
\begin{enumerate*}
	\item \emph{per-state resolution}, where hazards on each variable can be resolved separately,
	\item \emph{resolution policy}, where each access can be configured to either stall until hazards clear or speculatively proceed and recover later,
	\item \emph{per-address resolution}, where hazards on indexed state can be precisely resolved at address granularity, and
	\item \emph{selectable observation stage}, where designers choose the pipeline stage at which each variable is read.
\end{enumerate*}
However, the model's concurrent semantics place the burden on designers to ensure that their synchronization logic implements the intended hazard resolution strategy both correctly and efficiently.

PDL~\cite{pdl,xpdl} exposes control over pipelining strategy through predefined hazard resolution modules, but suffers from fragile, rule-dependent sequential semantics, partial control, and inefficient synchronization logic, as detailed in \cref{background:pdl-limitations}.
\begin{enumerate*}
	\item
	      PDL guarantees sequential semantics only if the program follows eighteen cumbersome rules on its use of the predefined modules, a burden aggravated by the many path-sensitive rules, which force designers to rewrite branch conditions into forms the compiler can analyze.
	      Moreover, we found that this guarantee is flawed, as some rule-compliant programs compile to pipelines that violate sequential semantics.
	\item PDL offers only partial control over hazard resolution along the four dimensions of \cref{intro:tab:granularity}, since it provides separate APIs for shared state (the hazard-lock API) and speculation.
	\item The predefined hazard resolution modules cannot exploit pipeline characteristics to optimize synchronization logic.
\end{enumerate*}
The latter two limitations make PDL-generated CPUs run $1.2\text{--}1.8\times$ slower and consume $1.8\text{--}4.1\times$ more area than RTL designs~(\cref{evaluation:riscv}).

\parhead{Our approach}
We present a HLS tool that delivers all three properties at once~(\cref{intro:tab:hls_comparison}):
inside sequential programming model, we
provide fine-grained control over pipeline strategies
and generates pipeline-optimized synchronization logic implementing the strategies.
We make the following contributions.

In \cref{background:visibility}, we introduce \emph{visibility control}, a novel programming abstraction in which controlling hazard resolution reduces to two choices: what an earlier instruction exposes (\ie, makes visible) of its pending reads and writes on shared state, and how a later instruction reacts to what is visible.
To stall, for example, the earlier instruction that writes to the shared state (the writer) delays exposing its value, and the later instruction that reads it (the reader) waits until the value becomes visible.
To bypass, the writer exposes its value early, and the reader uses it immediately.

In \cref{overview}, we instantiate visibility control in a sequential programming model that gives fine-grained control over the pipelining strategy.
We base our model on speculative loop pipelining, where each iteration overlaps with subsequent iterations~(\cref{overview:basic}).
We resolve dependency hazards via \code{Var}, a unified primitive for visibility control~(\cref{overview:var}).
We capture \emph{advisory state} (\eg, a branch predictor), which affects performance but not correctness, via \code{SpecVar}, a primitive that needs no hazard resolution~(\cref{specvar}).

In \cref{rtl-compilation}, we show how the compiler translates the chosen pipelining strategy into optimized RTL.
It factors the program-order comparison between in-flight instructions into one mechanism shared by every \code{Var}.
It optimizes each \code{Var}'s implementation for the pipeline's potential hazards, generating only the necessary registers and combinational logic.

In \cref{evaluation}, we show across three case studies that our approach achieves competitive PPA.
Our designs outperform those from sequential-semantics HLS tools (PDL, Vitis HLS, Dynamatic), and match or exceed hand-written RTL and Kanagawa designs, both built under concurrent semantics.
On in-order RISC-V CPUs targeting a 7nm ASIC flow~\cite{asap7}, our designs achieve $1.2\text{--}1.8\times$ lower runtime and $1.7\text{--}3.9\times$ smaller area than PDL's.
They match hand-written RTL in cycle count and maximum frequency, staying within $6\%$ in area and $4\%$ in power.
On countif histograms, our designs match or exceed Kanagawa's PPA across static, dynamic, and speculative hazard resolution strategies.
On an AES accelerator, viewing the design as a sequential program surfaces two optimizations that reduce area over an RTL reference.

In the supplementary material~\cite{supp}, we provide the appendix and artifact.
The appendix presents
\begin{enumerate*}
	\item PDL's eighteen rules and the unsoundness evidence (\cref{appendix:pdl_comparison}), and
	\item extended RISC-V evaluation (\cref{appendix:riscv-eval}).
\end{enumerate*}
The artifact (\cref{appendix:sec:ae}) contains
\begin{enumerate*}
	\item the compiler source code,
	\item the designs of the three case studies,
	\item a token-trace visualizer for performance debugging, and
	\item evaluation scripts.
\end{enumerate*}

}
{\section{Background and Motivation}
\label{background}

We review dependency hazards and their resolution strategies (\cref{background:cpu-hazards}) and examine how PDL~\cite{pdl}, the closest existing approach to our goals, implements a 5-stage CPU (\cref{background:token-model}).
We then identify three shortcomings that motivate our design: a fragile guarantee of sequential semantics, partial control over resolution strategies, and inefficient strategy implementation (\cref{background:pdl-limitations}).

\subsection{Dependency Hazards}
\label{background:cpu-hazards}

\emph{Dependency hazards} arise when pipelined execution reorders operations on shared state, \eg, the register file, program counter (PC), or control and status registers (CSRs) of a CPU.
They violate sequential semantics in three forms:
Read-After-Write (RAW) reads state before an earlier write commits,
Write-After-Write (WAW) commits a write before an earlier one, leaving a stale value,
and Write-After-Read (WAR) commits a write before an earlier read finishes.

The simplest resolution strategy is to \emph{stall} the pipeline until the hazard clears.
However, stalling reduces throughput, motivating more aggressive strategies.

RAW hazards can be resolved by:
\begin{enumerate*}
	\item \emph{bypassing}, where the writer forwards its result to the reader before committing, at the cost of additional multiplexers and wiring, and
	\item \emph{speculation}, where the reader proceeds with a predicted value and recovers if the prediction is wrong, at the cost of recovery logic.
\end{enumerate*}
A representative application of speculation is resolving CPU control hazards (RAW on the PC) by predicting the next PC and flushing the pipeline on misprediction.

WAW and WAR hazards can be resolved by:
\begin{enumerate*}
	\item \emph{deferred commit}, where writes are delayed to a later stage, so writes commit in program order (WAW) and all prior reads complete (WAR), at the cost of pipeline registers, and
	\item \emph{register renaming}, where each logical write maps to a fresh physical location, removing the name dependency, at the cost of a rename table and free list.
\end{enumerate*}

\subsection{PDL: Pipeline Structure and Hazard Resolution}
\label{background:token-model}

PDL~\cite{pdl} exposes control over hazard resolution through two kinds of external subsystems, hazard locks and speculation, each implemented as a predefined blackbox module.
Its follow-up XPDL~\cite{xpdl} adds exceptions as another subsystem.
The designer writes the datapath pipeline and sends requests to the subsystems from its stages.
\Cref{fig:pdl-overview} shows a 5-stage pipelined CPU written in PDL.
\Cref{fig:pdl-cpu} is the program describing its datapath pipeline, and \cref{fig:pdl-model} shows how the compiler connects the datapath pipeline to the subsystems.

\begin{figure}[t]
	\centering
	\begin{subfigure}[c]{0.47\linewidth}
		\centering
\begin{minted}{rust}
pipe cpu(pc)[rf, imem, dmem, csrs, bht]{
  spec_check();
  acquire(imem[pc], R); insn <- imem[pc]; release(imem[pc]);
  --- // Fetch -> Decode
  spec_check();
  s <- spec_call cpu(pc + (bht.req(pc) ? imm : 4));
  acquire(rf[rs1], R); // rs2 omitted
  alu_arg1= rf[rs1];
  release(rf[rs1]);
  reserve(rf[rd], W);
  --- // Decode -> Execute
  spec_barrier();
  alu_out= alu(args*); // compute logic
  npc = calc_npc(args*); // branch target
  if (isIllegal(op)) { throw(ERR_INV); }
  if (npc== pc+4){verify(s, npc) { bht.upd(pc, brTaken) }}
  else {invalidate(s); call cpu(npc);}
  --- // Execute -> MemOps
  acquire(dmem[alu_out]);
  if (isStore(op)) {dmem[alu_out] <- data;}
  if (isLoad(op)) {dmem_out <- dmem[alu_out];}
  else {dmem_out = alu_out;}
  release(dmem[alu_out]);
  --- // MemOps -> Writeback
  block(rf[rd]); rf[rd] <- dmem_out;
  commit: release(rf[rd]);
  except(error_code):
    abort(rf[rd]); csrs[mepc] <- pc;
    csrs[mcause] <- error_code;
    call cpu(HANDLER_ADDR);
}
\end{minted}
		\caption{Abbreviated 5-stage pipelined CPU in PDL.}
		\label{fig:pdl-cpu}
	\end{subfigure}
	\hfill
	\begin{subfigure}[c]{0.52\linewidth}
		\centering
		\includesvg[width=0.65\linewidth]{fig/PDL-model.svg}
		\caption{Datapath pipeline generated from (a), connected to the subsystems PDL provides.}
		\label{fig:pdl-model}
	\end{subfigure}
	\caption{PDL's programming model: the datapath program (a) and the resulting CPU structure (b).}
	\label{fig:pdl-overview}
\end{figure}

\parhead{Datapath pipeline}
The datapath pipeline is the part the designer writes directly, by
\begin{enumerate*}
	\item designating the pipeline structure,
	\item writing the combinational logic of each stage, and
	\item sending requests to the subsystems.
\end{enumerate*}
It is written as a recursive function \code{cpu} (\cref{fig:pdl-cpu}), parameterized by the program counter (\code{pc}) and the shared resources it accesses: the register file (\code{rf}), instruction/data memory (\code{imem}/\code{dmem}), CSRs (\code{csrs}), and the branch history table (\code{bht}).
The pipeline structure is designated by two constructs.
The stage separators (\code{-{}-{}-}, L4, 11, 18, 24) mark the boundaries at which the compiler inserts pipeline registers.
The recursive \code{call cpu(npc)} (L17) spawns a new instruction that runs concurrently from L2, overlapping the spawner's remaining stages.
Statements in a single stage are combinational, \eg, computing the ALU result and the branch target (L13--14).
The remaining statements are requests to the subsystems, such as the register file read request (L7).
We describe the three kinds of subsystems in turn: hazard locks (guarding memories such as \code{rf} and \code{dmem}), speculation~(drawing predictions from \code{bht}), and exceptions~(manipulating \code{csrs}).

\parhead{Hazard locks}
PDL resolves dependency hazards as resource contention using \emph{hazard locks}.
A lock tracks accesses at address granularity within its memory, \eg, individual registers in the register file.
Each access follows a three-operation protocol:
\begin{enumerate*}
	\item \code{reserve} records access intent in program order,
	\item \code{block} stalls until the access is granted, after which a read or write executes, and
	\item \code{release} commits the result to the shared state and frees the lock.
\end{enumerate*}
For a given address, \code{block} grants access in \code{reserve} order.
PDL provides different kinds of predefined lock modules for distinct resolution strategies: Queue Lock (stall), Bypass Queue (bypassing), and Renaming Register File (register renaming).

The protocol resolves a RAW hazard by serializing the concurrent accesses.
For example, consider two back-to-back instructions where the first writes \code{r1} and the second reads \code{r1}.
The read lock is \code{acquire}d in \code{Decode} (L7), where \code{acquire} is a shorthand combining \code{reserve} and \code{block} in the same stage.
The write lock access is split across stages:
\code{reserve} (L10) records the write intent in \code{Decode},
\code{block} (L25) defers the actual stall to \code{Writeback}, and \code{release} (L26) commits the write.
Since the first instruction \code{reserve}s its write before the second \code{acquire}s its read, the \code{acquire} stalls until the first instruction's access to the address is finished.

\parhead{Speculation}
PDL provides a separate \emph{speculation API} that coordinates instructions through a \emph{speculation table}, an automatically instantiated shared resource tracking in-flight instructions' speculative status.
An instruction can spawn a speculative successor with a predicted PC using \code{spec\_call} (L6), creating an entry in the speculation table.
The prediction comes from an integrated RTL branch predictor, queried by \code{bht.req} (L6) for whether the branch is taken.
In \code{Execute}, the branch is resolved and the spawner writes the result to the table.
\code{verify} (L16) compares the predicted and the resolved value, marking the entry as killed on mismatch or committed otherwise, and \code{bht.upd} (L16) updates the predictor with the outcome.
\code{invalidate} and \code{call} (L17) kill the entry and spawn the correct successor.

The spawned instruction, in turn, reads the speculation table to learn whether it is killed.
\code{spec\_check} (L2, 5) terminates the instruction without blocking if its entry is killed, gating lock reservation and further speculative calls.
\code{spec\_barrier} (L12) blocks until the entry resolves, then terminates the instruction if killed, gating write-lock release.
Both gates prevent a killed instruction from corrupting locks and the speculation table.

\parhead{Exceptions}
Exceptions are handled in two \emph{final blocks} that follow the pipeline body.
The \code{commit} block (L26) executes on the normal path, releasing the write lock.
The \code{except} block (L27--30) handles the exceptional path, calling \code{abort} (L28) to release the lock without committing, saving exception state to CSRs (L28--29), and redirecting to the exception handler via \code{call} (L30).

\subsection{Limitations of PDL}
\label{background:pdl-limitations}

PDL's reliance on separate subsystems for hazard resolution limits both usability and PPA.
In terms of usability, the cumbersome and unsound rules required for using the subsystems make PDL's guarantee of sequential semantics fragile.
In terms of PPA, the subsystems offer only partial control over resolution strategies and implement them without exploiting pipeline characteristics, making PDL-generated CPUs run $1.2\text{--}1.8\times$ slower and consume $1.8\text{--}4.1\times$ more area than RTL designs~(\cref{evaluation:riscv}).

\parhead{Fragile guarantee of sequential semantics}
PDL guarantees sequential semantics only if the compiler can prove that the datapath pipeline follows the rules each subsystem's API assumes, and the rules are both cumbersome and unsound.
\begin{enumerate*}
	\item The rules are cumbersome~(\cref{appendix:pdl:sequential_semantics}).
	      PDL and XPDL impose eighteen rules, including both per-subsystem and cross-subsystem rules, and many of them are path-sensitive.
	      Path-sensitive rules (\eg, on every program path, a hazard lock must be reserved, blocked, and released in order) require designers to reason about the branch conditions under which each operation can execute and to rewrite them into forms the compiler can analyze.
	      Cross-subsystem rules (\eg, speculative instructions must never release write locks) require designers to coordinate requests across subsystems.
	\item The rules are unsound~(\cref{unsoundness}).
	      Some rule-compliant programs still compile to pipelines that violate sequential semantics.
	      For instance, two conditional \code{release} operations placed in different pipeline stages can commit out of order, producing an unresolved WAW hazard.
\end{enumerate*}

\parhead{Partial control over resolution strategies}
PDL provides hazard locks and \code{spec\_call} as separate mechanisms, which prevents it from fully supporting the four dimensions of \cref{intro:tab:granularity}:
\begin{enumerate*}
	\item \emph{per-state resolution}: \code{spec\_call} bundles all speculative arguments into an atomically verified entry, so a misprediction on one argument forces recovery on all others,
	\item \emph{resolution policy}: each state uses only one mechanism, so its accesses cannot mix conservative and speculative resolution,
	\item \emph{per-address resolution}: \code{spec\_call} cannot express indexing, so PDL speculation cannot distinguish addresses, and
	\item \emph{selectable observation stage}: \code{spec\_call(value)} takes the prediction as a call argument, so PDL pins the predicted value at spawn time.
\end{enumerate*}

\parhead{Inefficient synchronization logic}
PDL builds each subsystem separately, unaware of the others and of the pipeline structure, so the subsystems can neither share synchronization logic nor specialize it to the pipeline.
For example, the CPU of \cref{fig:pdl-cpu} incurs two overheads:
\begin{enumerate*}
	\item redundant program-order tracking, since the three hazard locks and the speculation subsystem each keep separate per-instruction metadata for program order, carried in pipeline registers, and
	\item unnecessary order comparison, since the subsystems compare program order at runtime even when the pipeline structure determines it statically, \eg, an instruction in \code{Writeback} necessarily precedes one in \code{Decode}.
\end{enumerate*}
These overheads preclude using hazard locks even for simple stall-only state, forcing XPDL to implement CSRs via exceptions.\footnote{XPDL mentions this overhead as ``using locks to guard CSRs would be expensive''~\cite[\S4.1]{xpdl}.}

}
{\section{Hazard Resolution as Visibility Control}
\label{background:visibility}

We unify the hazard resolution strategies of \cref{background:cpu-hazards} in a programming abstraction called \emph{visibility control}.
We proceed in two steps:
\begin{enumerate*}
	\item[(\cref{background:visibility:examples})] building intuition with three example pipelines whose instructions resolve hazards by controlling when pending effects become \emph{visible} to later instructions, and
	\item[(\cref{background:visibility:abstraction})] defining the abstraction and specifying example strategies with its small set of operations.
\end{enumerate*}
Visibility control targets dependency hazards within a single program order.\footnote{Coordinating operations that do not share a program order, for example to enforce memory consistency across cores, remains future work.}
Later we implement stall, bypass, speculation, deferred commit, and register renaming using this abstraction (\cref{overview:var}).

\subsection{Motivating Examples}
\label{background:visibility:examples}

The three example pipelines (\cref{data_hazard:fig:cpu-behavior}) resolve hazards by deciding
\begin{enumerate*}
	\item what the earlier instruction exposes of its pending reads and writes on shared state and
	\item how the later instruction reacts.
\end{enumerate*}
Each panel's writer reveals its pending effect in steps to a reader acting on what is visible so far.

The first two pipelines resolve the same RAW hazard on the register file, by bypassing (\cref{fig:cpu-behavior-raw-conservative-bypass}) and by stalling (\cref{fig:cpu-behavior-raw-conservative-stall}).
In both, \tokenlabel{LD} stores to \code{r1} and the immediately following \tokenlabel{ADD} loads it.
In the bypass pipeline, \tokenlabel{LD} reveals its store incrementally:
\begin{enumerate*}
	\item at \code{Decode} it announces that it will write \code{r1},
	\item at \code{Memory} it makes the value visible,
	\item and at \code{Writeback} it commits the value to the register file.
\end{enumerate*}
\tokenlabel{ADD} reacts to each step.
It stalls while only the target address is known, and forwards the value the moment it becomes visible.
In the stall pipeline, \tokenlabel{LD} never makes its value visible early, so \tokenlabel{ADD} waits until the register file is written.
Waiting costs cycles but eliminates the bypass network (the multiplexers and wiring that forwarding requires).

The third pipeline resolves the control hazard on the program counter by speculation (\cref{fig:cpu-behavior-control-spec}).
The reader proceeds with whatever \code{pc} is visible, even though the value may still change.
\tokenlabel{XOR} is therefore fetched with \tokenlabel{BEQ}'s predicted \code{pc} and is flushed when \tokenlabel{BEQ} replaces the prediction.

\begin{figure}[t]
	\centering

	\begin{subfigure}[b]{0.58\linewidth}
		\centering
		\includesvg[width=\linewidth]{fig/section3-load-use-bypass.svg}
		\caption{Conservative hazard resolution on the register file (bypass).}
		\label{fig:cpu-behavior-raw-conservative-bypass}
	\end{subfigure}\hfill
	\begin{subfigure}[b]{0.30\linewidth}
\begin{minted}{rust}
// Decode
rs = load rf[rs_addr]
decl rf[rd_addr]
// Execute
if exe_wb then
  store wb_data to rf[rd_addr]
  seal rf[rd_addr]
// Memory
if mem_wb then
  store wb_data to rf[rd_addr]
  seal rf[rd_addr]
// Writeback
commit rf[rd_addr]
\end{minted}
		\caption{Specification of (\subref{fig:cpu-behavior-raw-conservative-bypass}).}
		\label{fig:cpu-behavior-raw-conservative-bypass-code}
	\end{subfigure}

	\begin{subfigure}[b]{0.58\linewidth}
		\centering
		\includesvg[width=\linewidth]{fig/section3-load-use-stall.svg}
		\caption{Conservative hazard resolution on the register file (stall).}
		\label{fig:cpu-behavior-raw-conservative-stall}
	\end{subfigure}\hfill
	\begin{subfigure}[b]{0.30\linewidth}
\begin{minted}{rust}
// Decode
rs = load rf[rs_addr]
decl rf[rd_addr]
// Execute
if exe_wb then
  store wb_data to rf[rd_addr]
// Memory
if mem_wb then
  store wb_data to rf[rd_addr]
// Writeback
commit rf[rd_addr]
\end{minted}
		\caption{Specification of (\subref{fig:cpu-behavior-raw-conservative-stall}).}
		\label{fig:cpu-behavior-raw-conservative-stall-code}
	\end{subfigure}

	\begin{subfigure}[b]{0.58\linewidth}
		\centering
		\includesvg[width=0.96\linewidth]{fig/section3-control2.svg}
		\caption{Speculative hazard resolution on the program counter.}
		\label{fig:cpu-behavior-control-spec}
	\end{subfigure}\hfill
	\begin{subfigure}[b]{0.30\linewidth}
\begin{minted}{rust}
// Fetch
curr_pc = spec_load pc
decl pc
// Decode
store (curr_pc + 4) to pc
// Execute
if br_taken then
  store next_pc to pc
\end{minted}
		\caption{Specification of (\subref{fig:cpu-behavior-control-spec}).}
		\label{fig:cpu-behavior-control-spec-code}
	\end{subfigure}

	\caption{Dependency hazard resolution in pipelined CPUs as visibility control.
Boolean flags \code{exe\_wb} (\code{Execute}) and \code{mem\_wb} (\code{Memory}) show whether an instruction produces its register-writeback value at each named stage.}
	\label{data_hazard:fig:cpu-behavior}
\end{figure}

\begin{figure*}[t]
	\centering
	\scriptsize

	\begin{subfigure}[b]{0.50\linewidth}
		\centering
		\small
		\[
		\resizebox{\linewidth}{!}{$
			\begin{array}{rcl}
				\sdecl{}  & ::= & \maydeclare \mid \nodeclare                                   \\
				\sstore{} & ::= & \bot \mid \declared(a) \mid \tentative(a,v) \mid \sealed(a,v) \\
				\sload{}  & ::= & \mayload \mid \noload
			\end{array}
		$}
		\]
		\caption{State definitions of \sdecl{}, \sstore{}, and \sload{}, with address $a$ and value $v$.}
		\label{fig:visibility-formal-states}
	\end{subfigure}
	\hfill
	\begin{subfigure}[b]{0.48\linewidth}
		\centering
		\scriptsize
		\includesvg[width=\linewidth]{fig/store_stm.svg}
		\caption{State machines \sdecl{}, \sstore{}, and \sload{}.}
		\label{fig:store-state-machine}
	\end{subfigure}

	\vspace{1.5em}

	\begin{subtable}[b]{0.60\linewidth}
		\scriptsize
		\centering
		\begin{tabular}{|l|c|c|c|}
			\hline
			$(\sdecl, \sstore{})$
			                           & \api{load} & \api{spec\_load} & \api{commit} \\
			\hline
			$(\maydeclare, \_)$        & \Stall     & \MemAccess       & \Stall       \\
			\hline
			$(\nodeclare, \declared)$  & \AddrStall & \MemAccess       & \AddrStall   \\
			\hline
			$(\nodeclare, \tentative)$ & \AddrStall & \Bypass          & \AddrStall   \\
			\hline
			$(\nodeclare, \sealed)$    & \Bypass    & \Bypass          & \AddrStall   \\
			\hline
			$(\nodeclare,\; \bot)$     & \MemAccess & \MemAccess       & \MemAccess   \\
			\hline
		\end{tabular}
		\caption{Action given $(\sdecl, \sstore)$.}
		\label{tab:visibility-control-store}
	\end{subtable}
	\hfill
	\begin{subtable}[b]{0.34\linewidth}
		\scriptsize
		\centering
		\begin{tabular}{|c|c|}
			\hline
			$\sload$
			           & \api{commit} \\
			\hline
			$\mayload$ & \Stall       \\
			\hline
			$\noload$  & \MemAccess   \\
			\hline
		\end{tabular}
		\caption{Action given $\sload$.}
		\label{tab:visibility-control-load}
	\end{subtable}

	\caption{
		Both tables name the current instruction's operations in columns and specify its corresponding responses in entries, based on an earlier instruction's published state.
		The row tuple $(\sdecl, \sstore)$ of (c) records whether the earlier instruction may still declare a store ($\sdecl$) and its pending-store state ($\sstore$).
		In (d), $\sload$ indicates whether the earlier instruction may still load.
		Symbols: \Stall{} (stall unconditionally), \AddrStall{} (stall if address conflicts), \Bypass{} (use value if address conflicts), \MemAccess{} (read from or write to the base state), and $\_$ (any state).
	}
	\label{tab:visibility-control}
\end{figure*}

\subsection{Visibility Control as a Programming Abstraction}
\label{background:visibility:abstraction}

Visibility control models \emph{shared state} loaded and stored by in-flight instructions, with operations for each to \emph{publish} its pending effects and \emph{observe} those of earlier instructions.
The shared state splits into two parts.
The \emph{base state} is underlying storage that instructions ultimately commit to, \eg, the register file and program counter.\footnote{A base state can be any stateful element, such as registers or SRAMs.}
Per-instruction \emph{published states} expose pending stores and loads to later instructions, as \tokenlabel{LD} did incrementally (\cref{fig:cpu-behavior-raw-conservative-bypass}).

\parhead{Published states}
On each shared state, an instruction's published states consist of three state machines (\cref{fig:store-state-machine}): two for stores (\sdecl{} and \sstore{}) and one for loads (\sload{}).
For stores, each instruction resolves its store incrementally, first determining \emph{whether} it stores (\sdecl{}), then \emph{where}, then \emph{what} (both \sstore{}):
\begin{enumerate*}
	\item \sdecl{} indicates whether the instruction may initiate a new store lifecycle ($\maydeclare$),
	      transitioning to $\nodeclare$ once the instruction is known not to do so.
	\item \sstore{} tracks a pending store through four states, advanced by three \emph{publish operations}.
	      \api{decl} resolves the address, advancing $\bot$ (no store declared) to $\declared(a)$.
	      \api{store} exposes a value that may still change, producing $\tentative(a, v)$.
	      \api{seal} finalizes the value as $\sealed(a, v)$.
	      When the store is written to the base state, \sstore{} returns to $\bot$.
\end{enumerate*}
These states name what \tokenlabel{LD} revealed in \cref{fig:cpu-behavior-raw-conservative-bypass}.
\tokenlabel{LD} declared \code{r1} at \code{Decode} and sealed the value at \code{Memory}, whereas the stall and speculation panels leave values \tentative{}.

For loads, \sload{} records whether the instruction may perform a load ($\mayload$), transitioning to $\noload$ once the instruction is known not to load.

\parhead{Visibility-driven operations}
An instruction acts on the published states of earlier instructions through three operations, and \cref{tab:visibility-control} specifies the action each operation takes given those states.

A \emph{conservative load} (\api{load}) reads only a finalized value, resolving RAW hazards through its column of \cref{tab:visibility-control-store}.
As long as an earlier instruction may still declare a store ($\maydeclare$), any address could be written, so the load stalls unconditionally (\Stall).
Once the address is known but the value is not final ($\declared$ or $\tentative$), the load stalls only on an address conflict (\AddrStall).
Once the value is $\sealed$, it can no longer change, so the load forwards it instead of stalling (\Bypass), the bypass of \cref{fig:cpu-behavior-raw-conservative-bypass}.
With no store pending ($\bot$), the load reads the base state (\MemAccess).

A \emph{speculative load} (\api{spec\_load}) reads eagerly, so its column contains no stall.
With no visible value ($\maydeclare$, $\declared$, or $\bot$), it reads the base state (\MemAccess).
Once a value is visible ($\tentative$ or $\sealed$), it forwards the value (\Bypass).
Any instruction that consumed a speculative value must be flushed when a preceding instruction's \sstore{} reaches or updates $\tentative$ at a conflicting address.
\tokenlabel{XOR} is flushed by this rule (\cref{fig:cpu-behavior-control-spec}).
The same mechanism can express exceptions, where the faulting instruction overwrites its \tentative{} \code{pc} with the exception handler's address.

A \api{commit} writes a finalized value to the base state, resolving WAW and WAR hazards through its columns in both tables.
For WAW, the commit stalls while an earlier instruction may still declare a store (\Stall) or already exposes one at a conflicting address (\AddrStall).
Stores to the same address therefore reach the base state in program order (\cref{tab:visibility-control-store}).
For WAR, the commit stalls while an earlier instruction may load ($\mayload$), so the earlier load still reads the old value (\cref{tab:visibility-control-load}).
Once every earlier instruction reaches $(\nodeclare, \bot, \noload)$, the write proceeds (\MemAccess).

\parhead{Specifying hazard resolution strategies}
With visibility control, we can specify hazard resolution strategies by selecting and placing publish and visibility-driven operations (\cref{data_hazard:fig:cpu-behavior}).
\cref{fig:cpu-behavior-raw-conservative-bypass-code} specifies the bypass behavior of \cref{fig:cpu-behavior-raw-conservative-bypass}.
The \api{decl} at L3 announces \tokenlabel{LD}'s destination \code{r1} at \code{Decode}.
The \api{store} and \api{seal} operations publish and finalize values at \code{Execute} (L6--7) or, for \tokenlabel{LD}, at \code{Memory} (L10--11).
The \api{load} at L2 stalls \tokenlabel{ADD} until \tokenlabel{LD}'s value is sealed, then forwards it (\cref{tab:visibility-control-store}).
The \api{commit} at L13 writes the value to the register file at \code{Writeback}.
\cref{fig:cpu-behavior-raw-conservative-stall-code} omits the two \api{seal} calls and specifies the stall behavior of \cref{fig:cpu-behavior-raw-conservative-stall}.
The \api{store} at L9 leaves \tokenlabel{LD}'s value \tentative{}, so \tokenlabel{ADD}'s \api{load} at L2 keeps stalling.
The load reads the register file after \api{commit} at L11 writes the value at \code{Writeback}.
The code in \cref{fig:cpu-behavior-control-spec-code} specifies the speculative behavior of \cref{fig:cpu-behavior-control-spec}.
The \api{decl} at L3 announces \tokenlabel{BEQ}'s pending store to \code{pc} at \code{Fetch}.
The \api{store} at L5 publishes the predicted \code{pc}.
The \api{spec\_load} at L2 lets \tokenlabel{XOR} consume that prediction at \code{Fetch}.
When the branch is taken (L7), \api{store} at L8 replaces the prediction with the corrected target at \code{Execute}, flushing \tokenlabel{XOR}.

}
{\section{Programming Dataflow Pipelines with Visibility Control}
\label{overview}

\noindent
We instantiate visibility control in a sequential programming model for dataflow pipelines (\cref{background:visibility}), letting designers easily control both pipelining pipeline structure and hazard resolution.

The model comprises three abstractions:
\begin{enumerate*}
\item[(\cref{overview:basic})] \emph{speculative loop pipelining} controls the pipeline structure, mapping each loop to a pipeline whose internal stages are set by PDL~\cite{pdl}'s separators;
\item[(\cref{overview:var})] the wrapper type \code{Var} controls hazard resolution per variable by instantiating visibility control; and
\item[(\cref{specvar})] the wrapper type \code{SpecVar} complements speculative loop pipelining and \code{Var} for advisory state, such as a branch history table (BHT) accessed freely from multiple stages.
\end{enumerate*}

\parhead{Source program}
\label{overview:source}
We realize the programming model in Rust, taking a program as specification and compiling it into a pipeline.
Although the proposed abstractions are language-agnostic, Rust offers two advantages: an advanced type system that enforces state-transition rules (\cref{fig:store-state-machine}), and rich testing frameworks for the source program.
We support a Rust subset defined by four constraints: safe Rust, single-threaded code over an acyclic call graph, no heap allocation, and every loop-carried variable wrapped in \code{Var} or \code{SpecVar}.
Within these constraints, arbitrary control flow, arbitrary bitwidth integer arithmetic, arrays, and stack access remain available.

\parhead{Target pipeline}
We target a dynamically scheduled dataflow pipeline whose stages each comprise a set of combinational operations.
Pipeline registers separate the stages, and a latency-insensitive protocol governs their execution~\cite{li-interface}.
We model work flowing through the stages with the standard \emph{token} abstraction of dataflow circuits~\cite{dynamatic}.
Each token represents one dynamic execution instance, carrying its in-flight values through pipeline registers in order.

\subsection{Speculative Loop Pipelining}
\label{overview:basic}

\begin{figure*}[tbp]
	\centering
	\footnotesize
	\begin{subfigure}[b]{0.31\linewidth}
		\centering
		\footnotesize
		\begin{minted}[linenos=true,breaklines, fontsize=\tiny, bgcolor=CodeBg, numbersep=2pt,xleftmargin=6pt]{rust}
// Marks stage boundaries.
fn sep() {} 
loop { // Loop head
  let (op, a, b) = ..; sep();
  let res = if op.is_add() { // Split
    let low = a[0..9] + b[0..9];
    sep();
    let high = a[9..] + b[9..] + low[9];
    sep();
    Some(high.append(low[0..9]))
  } else { None }; // Join
  sep();
  if res > THRESHOLD { break; }
}
\end{minted}
		\caption{Loop with \code{sep} and branch.}
		\label{fig:loop:source}
	\end{subfigure}
	\hfill
	\begin{subfigure}[b]{0.68\linewidth}
		\centering
		\includesvg[width=\textwidth]{fig/loop-structure.svg}
		\caption{Compiled pipeline from \cref{fig:loop:source}.}
		\label{fig:loop:execution}
	\end{subfigure}
	\caption{Compilation of a loop with a conditional two-stage adder.}
	\label{fig:loop}
\end{figure*}

\noindent
We compile each source loop into a pipeline that issues up to one token per cycle, each token executing one iteration.
\Cref{fig:loop:execution} illustrates the token flow of the pipeline compiled from \cref{fig:loop:source}.
When token \tokenlabel{P} reaches the loop head~(step~1), it remains there as the \emph{parent token} and issues one \emph{subtoken} per iteration~(step~2).
Subtokens inherit the \emph{program order} of their iterations: token \tokenlabel{$i$} \emph{precedes} token \tokenlabel{$j$} iff $i < j$.
(We generalize program order to nested loops in \cref{rtl-compilation:structure}.)

We adopt PDL's structural mapping to compile the loop body, making pipeline structure directly controllable through source changes.
The pipeline separators \code{sep()} (\cref{fig:loop:source}, L4, L7, L9, and L12) map to pipeline registers~(step~3),
and the \code{if} branch (L5--11) maps to a \emph{split} at its entry and a \emph{join} at its exit.
The split routes each token to the arm selected by the branch condition.
The join enforces \emph{in-order exit}: tokens leave in program order, so \tokenlabel{4} stalls until the earlier \tokenlabel{2} and \tokenlabel{3} exit~(step~4).

Loop pipelining is \emph{speculative} because token \tokenlabel{$i$} can start before source iteration $i$ is known to exist.
If an earlier token terminates the loop, the pipeline flushes all later tokens~(step~5) because they do not occur in source execution.
For safety, the compiler stalls each flushable token before any irreversible action.
After the flush, parent token \tokenlabel{P} leaves the loop head~(step~6).

\parhead{Comparison with explicit recursive calls}
Speculative loop pipelining differs from PDL's recursive calls in two respects:
\begin{enumerate*}
    \item Source iterations overlap without passing arguments.
    PDL expresses repeated execution through \code{call} with resolved arguments or \code{spec\_call} with predictions such as the next PC (\cref{background:token-model}).
    \item Source loops provide program order directly, whereas PDL requires each source path to specify one next instruction or terminate.
    This restriction establishes the instruction sequence on which all of PDL's hazard resolution mechanisms depend~\cite[\S3.2]{pdl}.
    Designers must therefore reason about branch conditions to exclude paths with multiple recursive \code{call}s, which would make the next instruction in program order ambiguous~\cite[\S4.3]{pdl}.
\end{enumerate*}

\subsection{Programming Visibility Control with \code{Var}}
\label{overview:var}

\begin{figure}[t]
	\centering
	\begin{subfigure}[b]{0.47\linewidth}
		\centering
\begin{minted}{rust}
/// Mutable array across loop iterations.
struct Var<T, const N: usize>([T; N]);
struct StoreBatch<T, const N: usize, const S: usize> {
  slots: [(usize, Option<T>); S],
  var_ref: &mut Var<T, N>,
};
struct Slot<const I: usize>;
enum Timing { Sync, Async };

impl Var<T, N> {
  fn load<Ti: Timing>(&self, addr: U<clog2(N)>)
     -> T { self.0[addr] }
  fn spec_load<Ti: Timing>(&self, addr: U<clog2(N)>) -> T { 
    self.0[addr] 
  }
  fn try_load(&self, addr: U<clog2(N)>) -> Option<T> {
    if random() { Some(self.0[addr]) } else { None }
  }
  fn prepare_batch(&mut self) -> StoreBatch<T, N, 0> {
    StoreBatch { slots: [], var_ref: self }
  }
}
impl StoreBatch<T, N, S> {
  fn decl(self, addr: usize) 
    -> (StoreBatch<T, N, S+1>, Slot<S>) {
    (StoreBatch {
      slots: self.slots.append((addr, None)),
      var_ref: self.var_ref
    }, Slot::<S>)
  }
  fn store<const I: usize>(&mut self, slot: &Slot<I>, v: T) {
    self.slots[I].1 = Some(v);
  }
}
impl Slot<I> {
  fn seal(self) {}
}
impl Drop for StoreBatch<T, N, S> {
  fn drop(&mut self) {
    for (addr, data) in self.slots {
      if let Some(v) = data { self.var_ref.0[addr] = v; }
    }
  }
}
\end{minted}
	\caption{Source-level API of \code{Var}.}
	\label{fig:source_semantics}
	\end{subfigure}
	\hfill
	\begin{subfigure}[b]{0.50\linewidth}
		\centering
\begin{minted}{rust}
fn core(
  init_pc: u32,
  imem: Fn(u32) -> MemResp,
  dmem: FnMut(MemReq) -> MemResp
) {
  let mut pc = Var::new([init_pc]);
  let mut rf = Var::new([0; 32]);
  loop { // Fetch stage
    let curr_pc = pc.spec_load::<Sync>(0);
    let inst = imem(curr_pc);
    let (mut npc_b, npc_slot) = pc.prepare_batch().decl(0);
    npc_b.store(&npc_slot, curr_pc + 4); // Predict PC as PC+4.
    sep(); // Decode stage
    let Decoded { rs1_addr, rs2_addr, rd_addr, exe_wb, .. }
     @ dec = decode(inst);
    let rs1 = rf.load::<Async>(rs1_addr); // rs2 omitted.
    let (mut rf_b, rf_slot) = rf.prepare_batch().decl(rd_addr);
    sep(); // Execute stage
    let (executed, next_pc, br_taken) = execute(dec, rs1, rs2);
    let memsel = if exe_wb {
      rf_b.store(&rf_slot, executed.wb_data);
      rf_slot.seal();
      MemSel::Exe
    } else { MemSel::Mem(rf_slot) };
    if br_taken { npc_b.store(&npc_slot, next_pc); }
    drop(npc_b);
    sep(); // Memory stage
    match memsel {
      MemSel::Exe => {},
      MemSel::Mem(rf_slot) => {
        let mem_res = memory(executed, &dmem);
        if let Some(wb_data) = mem_res.wb_data() {
          rf_b.store(&rf_slot, wb_data);
          rf_slot.seal();
        }
      }
    };
    sep(); // Writeback stage
    drop(rf_b);
  }
}
\end{minted}
		\caption{5-stage CPU pipeline with bypass and static prediction.}
		\label{fig:overview:cpu}
	\end{subfigure}
	\caption{Source-level API and CPU example of \code{Var}.}
	\label{fig:overview:var-cpu}
\end{figure}

\noindent
\code{Var} abstracts data with loop-carried dependencies, and its API instantiates visibility control~(\cref{background:visibility}).
The position of \code{Var}'s operations in the pipeline determines its hazard resolution strategy.

\parhead{Source behavior}
At the source level, \code{Var<T, N>} is an array \code{[T; N]} (\cref{fig:source_semantics}, L2).
\code{load} and \code{spec\_load} read an element, and \code{try\_load} randomly returns \code{Some} of that element or \code{None}.
\code{StoreBatch} writes its stored values to the array in declaration order when Rust's \code{Drop} commits it.
The publish operations of \cref{background:visibility:abstraction} prepare these writes in slots.
Each \code{decl} allocates a slot for an address, and \code{store} fills or updates its value.
Ownership types, typestates, and RAII enforce three usage rules, rejecting invalid operations at compile time:
\begin{enumerate*}
    \item A \code{StoreBatch} mutably borrows its \code{Var} (\cref{fig:source_semantics}, L5), so at most one batch is live for a given \code{Var} at a time.
    \item \code{seal} consumes the slot handle required by \code{store}, preventing further updates.
    \item Every batch commits exactly once.
\end{enumerate*}

\begin{figure}[tbp]
	\centering
	\begin{subfigure}[b]{0.45\linewidth}
		\centering
		\small
		\[
			\begin{array}{rcl}
				\sbatch{} & ::= & [\sstorei{1},\ldots, \sstorei{n}] \\
				\sspec{} & \subseteq & \{0,\ldots,N-1\} \\
			\end{array}
		\]
		\caption{State definitions of \sbatch{} and \sspec{}.}
		\label{fig:var-formal-states}
	\end{subfigure}
	\hfill
	\begin{subfigure}[b]{0.5\linewidth}
		\centering
		\includesvg[width=\linewidth]{fig/state-machine-compilation.svg}
		\caption{State transitions of \sbatch{} and \sspec{}.}
	\end{subfigure}
	\caption{States and transitions of \sbatch{} (list of independent \sstore{} slots) and \sspec{} (addresses \code{spec\_load} reads).}
	\label{fig:state-transition-machines}
\end{figure}

\parhead{Target behavior}
In the pipeline, each loop iteration's token plays the role of an instruction in visibility control (\cref{background:visibility}).
Tokens observe preceding tokens' published states in program order, with the committed array contents as the base state.
The conservative load, speculative load, and commit rules of \cref{tab:visibility-control} govern \code{load}, \code{spec\_load}, and \code{drop(batch)}, respectively.%
\footnote{For \code{load}, \code{Timing::Async} allows reading a sealed value within the same cycle, whereas \code{Timing::Sync} reads it on the next cycle (\cref{rtl-compilation:var}).}
The API adds \code{try\_load} as a nonblocking variant of the conservative load.
It immediately returns \code{None} when \code{load} would stall and \code{Some} of the loaded value otherwise.%
\footnote{The source specification permits either outcome so that the target can select a result according to pipeline hazards.}

The published state retains \sdecl{} and \sload{} while replacing the single pending \sstore{} with an ordered list \sbatch{} of write slots.
Each \code{decl} appends a \declared{} slot, whose subsequent \code{store} and \code{seal} operations follow the \sstore{} machine.%
\footnote{The source API statically restricts expressible batch histories to the transitions of \cref{fig:state-transition-machines}.}
\code{drop(batch)} waits until the inherited WAW and WAR checks pass for the whole batch and the token can no longer be flushed.
It then atomically writes the stored slots to the base state in declaration order.

Each token also records its speculative reads in an address set \sspec{}.
The set starts empty, and \code{spec\_load(addr)} inserts the address.
The speculative-load flush of \cref{background:visibility:abstraction} is triggered when a slot reaches or updates \tentative{} at an address recorded in a later token's \sspec{}.
The pipeline flushes that later token and all tokens after it in program order.%
\footnote{The flush shares the loop-termination mechanism of \cref{overview:basic}.}

\parhead{Example 1: 5-stage pipelined CPU}
\Cref{fig:overview:cpu} realizes the example pipelines of \cref{background:visibility} as \code{Var} code inside a speculative loop pipeline, combining two strategies: bypass on \code{rf} and speculation on \code{pc}.
For \code{rf}, the pipeline realizes the bypass example of \cref{fig:cpu-behavior-raw-conservative-bypass} and \cref{fig:cpu-behavior-raw-conservative-bypass-code}.
\code{decl(rd\_addr)} (\cref{fig:overview:cpu}, L17) opens a $\declared$ slot at \code{Decode}.
The \code{store} and \code{seal} calls advance the slot through $\tentative$ to $\sealed$ in \code{Execute} for ALU results (L21--22) or \code{Memory} for loads (L33--34).
A later token's \code{load} bypasses the sealed value in the same cycle (\cref{fig:overview:raw-bypass}).
The batch commits at \code{Writeback} (L39).
For \code{pc}, the pipeline realizes the speculation example of \cref{fig:cpu-behavior-control-spec} and \cref{fig:cpu-behavior-control-spec-code}.
\code{spec\_load} (\cref{fig:overview:cpu}, L9) reads \code{pc} and records the address in $\sspec{}$.
\code{store} (L12) publishes the prediction \code{curr\_pc}{+}4 as \tentative{} for the next token to fetch ahead.
A taken branch's \code{store} (L25) instead overwrites the \tentative{} value with \code{next\_pc}, invalidating the speculation and flushing every token that observed it (\cref{fig:overview:branch-flush}).
The batch commits at the end of \code{Execute} (L26).

\parhead{Design-space exploration}
Small edits to the CPU of \cref{fig:overview:cpu} change its hazard resolution strategy while preserving its sequential behavior.
Deleting \code{rf}'s \code{seal} calls (L22, L34) removes its bypass network and makes conflicting loads wait for commit, as in the stall pipeline of \cref{fig:cpu-behavior-raw-conservative-stall-code}.
Replacing \code{pc.spec\_load} with \code{pc.load} (L9) removes misprediction recovery but stalls fetch until the branch target resolves.
Moving \code{rf\_b}'s \code{drop} (L39) after its final use in Execute or Memory can reduce the pipeline registers carrying \sbatch{}, at the cost of additional WAW or WAR hazards detection logic.

\begin{figure}[tbp]
	\centering

	\begin{subfigure}[t]{\linewidth}
		\centering
		\includesvg[width=0.6\linewidth]{fig/alu-bypass.svg}
		\caption{
			\tokenlabel{SUB}'s \code{seal} land in the execute stage, so \tokenlabel{ADD}'s \code{load rs1} bypasses the sealed value in the same cycle.
		}
		\label{fig:overview:raw-bypass}
	\end{subfigure}

	\begin{subfigure}[t]{\linewidth}
		\centering
		\includesvg[width=0.56\linewidth]{fig/static-prediction-behavior.svg}
		\caption{
			\tokenlabel{BEQ}'s \code{store} on \code{pc} overwrites the \tentative{} value, expressing speculation invalidation.
		}
		\label{fig:overview:branch-flush}
	\end{subfigure}

	\caption{Hardware-module instantiation of \cref{data_hazard:fig:cpu-behavior} for \cref{fig:overview:cpu}.}
	\label{fig:overview:cpu-trace}
\end{figure}

\parhead{Hazard resolution expressiveness}
\code{Var} supports all four dimensions of hazard resolution (\cref{intro:tab:granularity}):
\begin{enumerate*}
\item each state can use its own strategy through a separate \code{Var} (per-state resolution),
\item one \code{Var} can mix conservative and speculative reads by choosing \code{load} or \code{spec\_load} per access, without changing loop structure (resolution policy),
\item speculative dependencies are tracked automatically per element within each \code{Var} by \sspec{} (per-address resolution), and
\item conservative and speculative reads can observe values after token issue at stages selected by their placement relative to \code{sep} (selectable observation stage).
\end{enumerate*}

\parhead{Example 2: renaming register file}
The CPU of \cref{fig:overview:rename} composes \code{Var}s to implement register renaming.
Three \code{Var}s replace the single \code{rf} (L1--3): a \emph{free list} of per-physical-register free bits (\code{free}), a \emph{rename table} mapping architectural to physical registers (\code{names}), and the physical register file (\code{phys}).
Reading an operand chains two loads (L8--9): a rename table lookup that bypasses the sealed mapping of an in-flight rename (L22), followed by a \code{phys} read.
The allocator scans the free list by calling \code{try\_load(i)} on each entry (L11).
Entries with in-flight writes return \code{None} and are skipped instead of stalling the scan, while a free register returns \code{Some(true)} and is selected (L10--12).
Once a physical register is allocated, the free list (L13--19) and the rename table (L21--22) are updated, and a slot for the pending write to \code{phys} is declared (L24).

\begin{figure}[tbp]
	\centering
	\begin{subfigure}[b]{0.57\linewidth}
		\centering
\begin{minted}{rust}
let mut free: Var<bool, PHYS_REGS> = ..; // Free list
let mut names: Var<PhysId, ARCH_REGS> = ..; // Rename table
let mut phys: Var<Word, PHYS_REGS> = ..; // Physical regfile
loop {
  .. // Fetch (same as before)
  sep(); // Decode / Rename stage
  ..
  let rs1_phys = names.load::<Sync>(rs1_addr);
  let rs1 = phys.load::<Async>(rs1_phys); // rs2 omitted.
  let rd_alloc =
    loop { if let Some(i) = (0..PHYS_REGS).map(|i| free.try_load(i))
      .find_idx(|f| f == Some(true)) { break i; } };
  let free_b = free.prepare_batch();
  // Free old physical register mapped to rd.
  let old_rd_phys = names.load::<Sync>(rd_addr);
  let (mut free_b, s1) = free_b.decl(old_rd_phys);
  free_b.store(&s1, true);
  // Allocate new physical register.
  let (mut free_b, s2) = free_b.decl(rd_alloc); free_b.store(&s2, false);
  // Update rename table.
  let (mut names_b, names_slot) = names.prepare_batch().decl(rd_addr);
  names_b.store(&names_slot, rd_alloc); names_slot.seal();
  // Prepare write to physical register.
  let (phys_b, phys_slot) = phys.prepare_batch().decl(rd_alloc);
  ..
}
\end{minted}
		\caption{Register renaming with \code{try\_load}.}
		\label{fig:overview:rename}
	\end{subfigure}
	\hfill
	\begin{subfigure}[b]{0.41\linewidth}
		\centering
		\begin{minted}[linenos,fontsize=\tiny, bgcolor=CodeBg, breaklines, numbersep=4pt]{rust}
let mut bht = SpecVar::new(Bht::new());
loop {
  let curr_pc = pc.spec_load::<Sync>(0);
  let inst = imem(curr_pc);
  let br_tgt = branch_target(inst, curr_pc);
  let (mut npc_b, npc_slot) =
    pc.prepare_batch().decl(0);
  let pred_tkn =
    predict_branch(bht.load(), curr_pc, inst);
  let pred_pc =
    if pred_tkn { br_tgt } else { curr_pc + 4 };
  npc_b.store(&npc_slot, pred_pc);
  ..
  let (executed, br_taken) =
    execute(decoded, rs1, rs2);
  if br_taken != pred_tkn {
    let correct_pc =
      if br_taken { br_tgt } else { curr_pc + 4 };
    npc_b.store(&npc_slot, correct_pc);
  }
  let bht_upd =
    bht.load().update(curr_pc, pred_tkn, br_taken);
  bht.store(bht_upd);
  ..
}
\end{minted}
		\caption{Dynamic prediction with \code{SpecVar}.}
		\label{fig:overview:bht}
	\end{subfigure}
	\caption{CPU extensions using \code{Var} and \code{SpecVar}.}
	\label{fig:overview:cpu-extensions}
\end{figure}

\subsection{Programming Advisory State with \code{SpecVar}}
\label{specvar}

We introduce \code{SpecVar<T>} for advisory state.
Its \code{load} returns an random value and its \code{store} is a no-op.
This deliberately weak contract lets us compile \code{SpecVar} to a shared storage that tokens access without hazard resolution.

\Cref{fig:overview:bht} illustrates dynamic branch prediction with a BHT using \code{SpecVar}.
Each token predicts whether the next branch will be taken based on the BHT state in \code{SpecVar} (L8--9) and updates it with the actual outcome (L21--23).
Correctness is independent of the arbitrary value returned by \code{bht.load()}.
If the prediction matches the outcome (\code{br\_taken == pred\_tkn}), the PC assigned at L12 is already correct.
Otherwise, the token overwrites it with \code{correct\_pc} (L17--19).

}
{\section{Compilation to RTL}
\label{rtl-compilation}

The compiler maps the pipeline structure and hazard resolution of \cref{overview} to synthesizable Verilog.
We implement the translation in a 38K-LOC Rust compiler plugin whose input is Rust's mid-level intermediate representation (MIR)~\cite{mir}.
We analyze the control-flow graph of MIR to identify loops and branches, then generate the RTL in three steps:
\begin{enumerate*}
	\item[(\cref{rtl-compilation:structure})] constructing the pipeline structure from \code{sep}, \code{loop}, and branches;
	\item[(\cref{rtl-compilation:var})] generating hazard-resolution networks for \code{Var}'s publish and visibility-driven operations; and
	\item[(\cref{rtl-compilation:cpu})] optimizing these circuits to obtain an efficient implementation.
\end{enumerate*}

\subsection{Compilation of Pipeline Structure}
\label{rtl-compilation:structure}

\parhead{Compilation of a token}

A token compiles to a bundle of signals with separate fields for its valid bit $\mathit{vld}$, published states $S$, \emph{loop-index list} $L$, and local variables $\mathit{data}$ (\cref{fig:rtl:sep-circuit}).
For every \code{Var} $v_i$ in scope, $S_{v_i}$ stores all four components of its published state, $(\sdecl{},\sbatch{},\sload{},\sspec{})$ (\cref{overview:var}).
$\sdecl{}$ and $\sload{}$ each occupy one bit.
For a \code{Var} with $N$ elements and $V$-bit values, $\sbatch{}$ concatenates its slot representations.
Each slot comprises a one-hot four-bit $\sstore{}$ discriminant, a $\lceil\log_2 N\rceil$-bit address, and a $V$-bit value.
$\sspec{}$ encodes the set of speculatively read addresses as an $N$-bit mask.
The loop-index list $L=[n_{\ell_1},\ldots,n_{\ell_m}]$ stores the encoded iteration numbers of the enclosing loops, ordered from outermost to innermost.
For a loop $\ell_j$ with at most $I_j$ simultaneous pipelined tokens, its index $n_{\ell_j}$ occupies $b_j=1+\lceil\log_2 I_j\rceil$ bits.
Finally, $\mathit{data}$ holds the token's necessary local variables.

For example, a token entering the Decode stage of the 5-stage CPU occupies 110 bits (\cref{fig:overview:cpu}).
Its fields are:
\begin{enumerate*}
	\item one valid bit,
	\item 39 bits for \code{pc}'s published state, comprising one bit for $\sdecl{}$, $4+0+32=36$ bits for its single-slot $\sbatch{}$, one bit for $\sload{}$, and one bit for $\sspec{}$,
	\item 34 bits for \code{rf}'s published state, comprising one bit for $\sdecl{}$, zero bits for its empty $\sbatch{}$, one bit for $\sload{}$, and 32 bits for $\sspec{}$,
	\item a four-bit loop index for up to five simultaneous tokens, one per stage, and
	\item 32 bits in $\mathit{data}$ for the local variable \code{inst}, which is used in later stages.
\end{enumerate*}

\parhead{Compilation of \code{sep}}

\begin{figure}[tbp]
	\centering
	{\phantomsubcaption\label{fig:rtl:sep}\label{fig:rtl:sep-source}\label{fig:rtl:sep-circuit}
		\phantomsubcaption\label{fig:rtl:loop}\label{fig:rtl:loop-source}\label{fig:rtl:loop-circuit}
		\phantomsubcaption\label{fig:rtl:branch}\label{fig:rtl:branch-source}\label{fig:rtl:branch-circuit}}%
	\includesvg[width=0.96\linewidth,inkscapelatex=false]{fig/pipeline-structure-rtl.drawio.svg}
	\caption{RTL translation of (a) \code{sep}, (b) \code{loop}, and (c) conditional branches.}
	\Description{Three wide panels integrate source fragments with RTL circuits. The sep panel shows a two-multiplexer elastic buffer, the loop panel shows a parent controller and separate continue and break circuits, and the branch panel shows a split and an ordered join. Red dashed lines denote external outputs, blue dotted lines denote external inputs, and black lines denote internal logic. Ready signals, metadata inputs, order predicates, and metadata width are shown explicitly. Plain fills denote combinational logic, and diagonal hatching denotes flip-flops.}
	\label{fig:rtl:pipeline-structure}
\end{figure}

A \code{sep} compiles to a one-token \emph{elastic buffer}~\cite{elastic-circuit-synthesis}, extended with a flush input and a metadata output (\cref{fig:rtl:sep-circuit}).
Its input and output valid-ready interfaces implement the latency-insensitive protocol of \cref{overview}, allowing a stalled token to remain stored across clock cycles.

A \code{sep}'s behavior depends on its stored token and three input signals.
Flip-flops hold the resident token $T_{\mathsf{sep}}$.
The input token $T_{\mathsf{in}}$ arrives from \code{stmts1}, and the downstream ready signal $\mathit{rdy}_{\mathsf{out}}$ indicates whether \code{stmts2} can accept the stored token.
The flush bit is the OR of all requests from \code{Var}s and enclosing loops to flush $T_{\mathsf{sep}}$.
External control prevents transfers of tokens selected for flushing at both interfaces.

The circuit computes three outputs from the stored token and the downstream ready signal.
The stored token directly drives $T_{\mathsf{out}}$, the output to \code{stmts2}.
The metadata output $\mathit{meta}_{\mathsf{sep}}$ supplies other control circuits with the valid bit, published states, and loop-index list.
The upstream ready signal $\mathit{rdy}_{\mathsf{in}}$ is asserted when \code{sep} is empty or $\mathit{rdy}_{\mathsf{out}}$ is high.

The next-state logic uses two multiplexers to either accept an incoming token, clear \code{sep}, or retain the stored token.
On an input transfer, the circuit accepts the incoming token by selecting $T_{\mathsf{in}}$.
Otherwise, an output transfer or flush clears \code{sep} by selecting $T_{\bot}$, an empty token with a cleared valid bit.
With no transfer or flush, the circuit retains the stored token by selecting $T_{\mathsf{sep}}$.

\parhead{Compilation of \code{loop}}

A \code{loop} compiles to two parts that implement the speculative loop pipelining of \cref{overview:basic} (\cref{fig:rtl:loop-circuit}):
\begin{enumerate*}
	\item a loop controller that stores the parent token and speculatively issues subtokens, and
	\item \code{continue} and \code{break} circuits that complete iterations in program order.
\end{enumerate*}

The loop controller comprises three parts (\cref{fig:rtl:loop-circuit}, left):
\begin{enumerate*}
	\item A \code{sep} stores the parent received as $T_{\mathsf{in}}$ from \code{stmts\_before} and provides a flush input and metadata output $\mathit{meta}_{l}$.
	\item The controller forms a subtoken $T_{\mathsf{issue}}$ by copying the parent's fields and appending a loop index to $L$ to encode program order.
	      The $b_j$-bit register $\mathit{ptr}$ supplies the next index for loop $\ell_j$, giving the list $\{L,\mathit{ptr}\}$.
	\item The issue interface transfers the subtoken when the parent is valid and \code{body}'s ready signal $\mathit{rdy}_{\mathsf{issue}}$ is asserted.
	      Whenever a subtoken is issued, $\mathit{ptr}$ increments for the next iteration.
\end{enumerate*}

The lexicographic order of loop-index lists determines the program order of tokens, denoted $\lpo$.
For two such lists $L$ and $L'$, $L\lpo L'$ holds exactly when
\begin{enumerate*}
	\item $L$ is a proper prefix of $L'$, so a parent precedes its subtokens (\eg, $[2]\lpo[2,1]$), or
	\item the lists first differ at their $j$-th entries, where $L[j]$ is the earlier index (\eg, $[2,2]\lpo[3]$).
\end{enumerate*}
Since loop $\ell_j$ holds at most $I_j$ tokens and its indices have $b_j=1+\lceil\log_2 I_j\rceil$ bits, $L[j]$ and $L'[j]$ differ by less than $2^{b_j-1}$, so $L[j]$ is the earlier index exactly when $L'[j]-L[j]$ has a zero most significant bit.
For example, with $I_j=4$ and $b_j=3$, iteration $8$ ($000_2$) follows iteration $7$ ($111_2$) despite its smaller encoding, since $000_2-111_2=001_2$.

The \code{continue} and \code{break} circuits use $\lpo$ to retire subtokens in program order, and \code{break} additionally requests loop termination from the controller (\cref{fig:rtl:loop-circuit}, right).
Each circuit checks the input token $T_{\mathsf{in}} = (\mathit{vld}, L)$ against the metadata $\mathit{meta}_i=(\mathit{vld}_i,L_i)$ of every other \code{sep} in the loop body.
The order check succeeds when $\mathit{vld}$ is asserted and, for every other \code{sep}, either $\mathit{vld}_i$ is cleared or $L\lpo L_i$ holds.
For \code{continue}, a successful check asserts the input ready signal $\mathit{rdy}_{\mathsf{in}}$, retiring the subtoken.
For \code{break}, success instead asserts the one-bit $\mathit{break}_l$ request to the controller's $\mathit{break}_l$ input.
The controller's AND gate combines the request with the downstream ready signal $\mathit{rdy}_{\mathsf{out}}$ from \code{stmts\_after} to enable the parent's output transfer.
The controller forwards $\mathit{rdy}_{\mathsf{out}}$ to \code{break} as $\mathit{rdy}_{\mathsf{break}}$, which an AND gate in \code{break} combines with the order check result to drive its input ready signal $\mathit{rdy}_{\mathsf{in}}$.
When the break token transfers, the circuit flushes the remaining tokens in the loop body.
The parent simultaneously transfers as $T_{\mathsf{out}}$ to \code{stmts\_after}.

\parhead{Compilation of \code{branch}}

A conditional branch compiles to a split and a join (\cref{fig:rtl:branch}).
The split asserts an arm's valid bit only when the input valid bit $\mathit{vld}_{\mathsf{in}}$ is set and the condition $x_{\mathsf{cond}}$ selects that arm.
The join uses $\lpo$ comparisons to enforce the in-order exit of \cref{overview:basic}.
It selects a valid output token $T_{\mathrm{arm}}$ or $T_{\mathrm{arm}'}$ as $T_{\mathsf{out}}$ only when that token precedes every valid token in the opposite arm.

\subsection{Compilation of \code{Var}}
\label{rtl-compilation:var}

\parhead{Compilation of publish operations}

\begin{figure}[tbp]
	\centering
	{\phantomsubcaption\label{fig:rtl:publish}
		\phantomsubcaption\label{fig:rtl:var-flush}
		\phantomsubcaption\label{fig:rtl:var-load}
		\phantomsubcaption\label{fig:rtl:var-commit}}%
	\includesvg[width=\linewidth,inkscapelatex=false]{fig/var-operations-rtl.drawio.svg}
	\caption{RTL circuits for \code{Var} operations.}
	\Description{Four panels show publish operations, the flush network, the load circuit, and the commit circuit. Blue dotted lines denote external inputs, red dashed lines denote external outputs, and black lines denote internal logic. The shared legend distinguishes plain combinational-logic fills from diagonally hatched flip-flops.}
	\label{fig:rtl:var-operations}
	\label{fig:rtl:var-publish-flush}
	\label{fig:rtl:var-visibility}
\end{figure}

A publish operation compiles to combinational logic over the token's published state (\cref{fig:rtl:publish}).
\code{decl}, \code{store}, and \code{seal} update \sbatch{} according to the slot transitions of \cref{overview:var}.
For an incoming batch $B=[s_0,\ldots,s_{n-1}]$, $\mathit{slot}_k$ identifies the encoded slot $B[k]$.
Each operation wires its updates as follows:
\begin{enumerate*}
	\item \code{batch.decl(addr)} concatenates $B$ with a new slot whose state bits encode \declared{} and whose address bits carry \code{addr},
	\item \code{batch.store(slot\_k, data)} sets the state bits of $B[k]$ to \tentative{} and connects \code{data} to its value bits,
	\item \code{slot\_k.seal()} sets the state bits of $B[k]$ to \sealed{}, and
	\item \code{spec\_load(addr)} adds its read address to $\sspec{}$ by ORing the incoming mask $M$ with the one-hot $N$-bit decoding $e_{\mathit{addr}}$ of \code{addr}.
\end{enumerate*}

\parhead{Compilation of flush network}

A \code{store} compiles to an additional flush network that implements the speculative-load flush of \cref{overview:var} (\cref{fig:rtl:var-flush}).
The network flushes the earliest mispredicted token and all later tokens in program order.
When a valid token executes \code{store}, it supplies $\mathit{store}=(\mathit{addr},L)$ with the slot's address and its own loop-index list.
The $\mathit{meta}_i=(\mathit{vld}_i,S_i,L_i)$ inputs come from every \code{sep} within the \code{Var}'s lifetime, where tokens may have read this \code{Var} speculatively.
The $\mathit{mayflush}_j=(\mathit{vld}_j,L_j)$ inputs carry the loop-index list from every \code{sep} in the circuit, since the network may also flush later tokens outside that lifetime.

The network comprises three parts:
\begin{enumerate*}
	\item A misprediction detector for each $\mathit{meta}_i$ flags a token whose speculative read conflicts with the store, asserting $\mathit{mispred}_i$ exactly when $\mathit{vld}_i$ is set, the store precedes it in program order ($L\lpo L_i$), and its speculation mask has the addressed bit $S_{i,\mathsf{spec}}[\mathit{addr}]$ set.
	\item An earliest misprediction finder ($\arg\min_{\lpo}$) selects the first mispredicted token in program order and outputs its loop-index list $L_{\mathsf{earliest}}$ and a bit $\mathit{mispred}_{\mathsf{earliest}}$ indicating whether any token is mispredicted.
	\item A flush decider for each $\mathit{mayflush}_j$ asserts $\mathit{flush}_j$ exactly when $\mathit{mispred}_{\mathsf{earliest}}$ and $\mathit{vld}_j$ are set and $L_j$ is at or after $L_{\mathsf{earliest}}$ in program order.
\end{enumerate*}

\parhead{Compilation of \code{load} and \code{spec\_load}}

A \code{load} or \code{spec\_load} compiles to a RAW-checking and value-selection circuit for \code{Var}'s visibility control (\cref{overview:var,fig:rtl:var-load}).
The request $\mathit{load}=(\mathit{is\_spec},\mathit{addr},L)$ contains a flag for \code{spec\_load}, the read address, and the loop-index list.
Every \code{sep} in the circuit supplies $\mathit{meta}_i=(\mathit{vld}_i,S_i,L_i)$, where $S_i$ is this \code{Var}'s published state.
For asynchronous loads, additional $\mathit{seal}_q$ inputs carry the same metadata after \code{seal}, making sealed values available for bypass in the same cycle.
The $\mathit{success}$ output bit indicates load completion, and $\mathit{data}$ carries the selected value.

The circuit comprises three parts:
\begin{enumerate*}
	\item A potential RAW hazard detector for each metadata or seal input $x$ asserts $\mathit{conflict}_x$ exactly when its token is valid, precedes the request ($L_x\lpo L$), and either may declare another store ($S_{x,\mathsf{decl}}=\maydeclare$) or has a slot at $\mathit{addr}$.
	\item A latest conflict finder first selects the nearest conflicting predecessor ($\arg\max_{\lpo}$) and outputs its full published state $S_{\mathsf{latest}}$ and a bit $\mathit{conflict}_{\mathsf{latest}}$ indicating whether any predecessor conflicts.
	      The finder then selects the last-declared slot $k^*$ at $\mathit{addr}$ in $S_{\mathsf{latest}}$'s batch and forms the reduced state $S'=(\sdecl{},\sstorei{k^*})$.
	      If no slot matches, the slot-state component is $\bot$.
	\item Action-decision logic looks up $\mathsf{Table}[S',\mathit{is\_spec}]$ in the conservative or speculative load column of \cref{tab:visibility-control-store}, treating a request with no conflicting predecessor ($\mathit{conflict}_{\mathsf{latest}}$ clear) as the row $(\nodeclare,\bot)$.
	      It clears $\mathit{success}$ for either stall action, and otherwise asserts $\mathit{success}$ while a multiplexer drives $\mathit{data}$ with slot $k^*$'s value for \Bypass{} or the value at $\mathit{addr}$ in base memory for \MemAccess{}.
\end{enumerate*}

\parhead{Compilation of \code{drop}}

A \code{drop(batch)} compiles to a commit circuit that implements the visibility-driven commit of \cref{overview:var} (\cref{fig:rtl:var-commit}).
The committing token can write base memory only when no predecessor poses a WAW or WAR hazard or may flush it.
The request $\mathit{commit}=(S_{\mathsf{batch}},L)$ carries the batch's slots and the committing token's loop-index list.
Every \code{sep} in the circuit supplies $\mathit{meta}_i=(\mathit{vld}_i,S_i,L_i)$ and $\mathit{mayflush}_j=(\mathit{vld}_j,S_j,L_j)$, where $S_j$ holds the published states of all \code{Var}s rather than only this one.
The $\mathit{success}$ output bit indicates whether the commit can proceed.

The circuit comprises three parts:
\begin{enumerate*}
	\item A potential WA(W$\mid$R) hazard detector for each $\mathit{meta}_i$ produces $\mathit{conflict}_i$ by ANDing $\mathit{vld}_i$, the order test $L_i\lpo L$, and the state test $\mathit{detect\_wawr}(S_i,S_{\mathsf{batch}})$.
	      The state test succeeds if the declaration component is \maydeclare{}, the load component is \mayload{}, or any pending slot's address matches a slot in $S_{\mathsf{batch}}$.
	\item A potential flush detector for each $\mathit{mayflush}_j$ asserts $\mathit{conflict}_j$ exactly when $\mathit{vld}_j$ is set, $L_j\lpo L$, and either the store test $\mathit{detect\_store}(S_j)$ or the break test $\mathit{detect\_break}(L_j,L)$ succeeds.
	      The store test checks whether any \code{Var} in $S_j$ has a \maydeclare{} declaration component or a still-writable \declared{} or \tentative{} slot.
	      The break test conservatively treats a predecessor that is not an ancestor of the committing token as capable of breaking a shared enclosing loop.
	\item Commit-decision logic ORs the conflict bits of both detectors and inverts the result to drive $\mathit{success}$ and base memory's $\mathit{write\_en}$ signal.
	      The stored slots in $S_{\mathsf{batch}}$ supply the addresses and values for the enabled writes.
\end{enumerate*}

\subsection{Optimized Pipeline Synchronization}
\label{rtl-compilation:cpu}

The compiler exploits pipeline structure and reachable published states to eliminate unnecessary comparisons, registers, and control logic (\cref{fig:rtl:cpu-opt}).

\parhead{Program-order analysis}

The compiler eliminates statically determined $\lpo$ comparisons and the registers that carry unused loop indices.
Two structural rules determine the order of tokens stored in \code{sep}s:
\begin{enumerate*}
	\item a loop controller's parent precedes all subtokens in its body, and
	\item a token precedes another if its \code{sep} is reachable from the other's along token flow, unless its \code{sep} lies in the body of the loop whose controller holds the other token.
\end{enumerate*}
When neither rule settles the full order, the compiler generates a comparator only for the innermost loop enclosing both \code{sep}s.
Each loop controller retains one parent and copies its loop-index list into every subtoken, so tokens within that loop share all outer indices.
The compiler then removes unused index fields from a \code{sep}'s registers, and a loop controller's counter once no \code{sep} in that loop keeps its index.

\parhead{State-reachability analysis}

The compiler uses reachability analysis on published states at each \code{sep} to reduce the published-state registers and optimize the control logic.
The analysis combines the effects of publish operations along incoming paths with whether subsequent control flow permits further \code{decl}s or \code{load}s.
For the resulting set of reachable states, the compiler generates only the registers needed to represent necessary $S$ and removes synchronization logic whose outcomes are statically known.

\begin{figure}[tbp]
	\centering
	\includesvg[width=\linewidth,inkscapelatex=false]{fig/cpu-opt-rtl.drawio.svg}
	\caption{Optimized RTL for the 5-stage CPU.}
	\Description{Three wide panels show the main pipeline, the register-file circuit, and the program-counter circuit. The main pipeline retains token flow, reverse ready signals, all four sep registers, and metadata, seal, store, and commit ports. The register-file circuit retains asynchronous seal bypasses, RAW checkers, the priority multiplexer, and load and commit results. The program-counter circuit retains its priority multiplexer, base register, constant fields, and flush fan-out. Blue dotted lines denote external inputs, red dashed lines denote external outputs, and black lines denote internal logic. Diagonal hatching distinguishes flip-flops from combinational logic.}
	\label{fig:rtl:cpu-opt}
\end{figure}

\parhead{Example: Generated 5-stage pipelined CPU}

The two analyses eliminate loop-index logic and reduce published-state registers across the CPU's main pipeline (\cref{fig:rtl:cpu-opt}).
Its \code{sep}s $s_1,\ldots,s_4$ form a single chain, fixing the order of valid body tokens from Writeback (earliest) to Fetch (latest).
The compiler therefore resolves every $\lpo$ comparison at compile time and removes loop indices from the \code{sep}s and $\mathit{ptr}$ from the loop controller.
The infinite loop makes \code{break} logic unnecessary.
State-reachability analysis fixes $\sdecl{}$, $\sload{}$, and $\sspec{}$ at each \code{sep}.
Therefore, only \code{rf}'s batch contents and \code{pc}'s pending values at $s_1$ and $s_2$ require published-state registers.
For example, the token entering Decode now occupies 65 bits instead of 110, comprising one valid bit $\mathit{vld}_d$, the 32-bit pending value $\mathit{npc}_d$, and the 32-bit local variable \code{inst} in $\mathit{data}_d$.

The compiler combines static program order with \code{rf}'s reachable published state $S$ to simplify its load, flush, and commit circuits:
\begin{enumerate*}
	\item For loads, the predecessors in Execute, Memory, and Writeback have $\sdecl{}=\nodeclare{}$ because Decode contains the iteration's only \code{rf} declaration.
	      The potential RAW hazard detectors therefore test only validity and equality with the requested source-register address $\mathit{rs\_addr}$.
	      The fixed order of those stages reduces the latest conflict finder to a priority multiplexer over these bits.
	      The slot state in $\sbatch{}$ still varies, so the conservative load retains a \sealed{} test to distinguish bypass from stall, with base memory as the fallback when no stage conflicts.
	      Asynchronous loading also retains the Execute and Memory seal multiplexers to bypass values sealed in the current cycle.
	\item For flushing, conservative loads keep $\sspec{}=\varnothing$, so the compiler removes the speculation-mask registers and the entire flush network.
	\item For commit, no instruction precedes Writeback, so the compiler removes WAW, WAR, and flush checks and fixes success to one.
\end{enumerate*}

The compiler optimizes \code{pc}'s load, flush, and commit circuits using its fixed program order and constant components of $S$:
\begin{enumerate*}
	\item For loads, all four \code{sep}s have $\sdecl{}=\nodeclare{}$, and only $s_1$ and $s_2$ retain \tentative{} \code{pc} slots.
	      Each $\mathit{conflict}$ bit therefore reduces to the valid bit, so the compiler removes the RAW hazard detectors and fixes load success to one.
	      Program order reduces the latest conflict finder to a priority multiplexer over the pending values $\mathit{npc}_d$ and $\mathit{npc}_e$, with the base register as the fallback.
	\item For flushing, tokens in these \code{sep}s have $\sspec{}=\{0\}$, and only Fetch and Decode can contain instructions later than an Execute store.
	      These constants remove the order and speculation-mask tests and the earliest misprediction finder.
	      The valid Execute \code{pc.store} signal directly drives $\mathit{flush}_{\mathit{dec}}$ and cancels the Fetch transfer.
	      The flush signal $\mathit{flush}_{\mathit{dec}}$ clears $s_1$ and suppresses its outgoing transfer.
	\item For commit, no predecessor can access \code{pc} or flush the token, so all checks disappear and the write always succeeds.
\end{enumerate*}

}
{\section{Evaluation}
\label{evaluation}

We evaluate our tool on three case studies chosen for two goals:
\begin{enumerate*}
  \item \emph{head-to-head comparison} with PDL and Kanagawa, the prior HLS tools that exploit ILP, and
  \item \emph{applicability} to control-heavy designs based on finite-state machines.
\end{enumerate*}
For head-to-head comparison, we reuse PDL and Kanagawa's workloads: \emph{RISC-V CPUs} (\cref{evaluation:riscv}) against both and \emph{countif histograms} (\cref{eval:histogram}) against Kanagawa.
For applicability, we implement an \emph{AES cipher accelerator} (\cref{eval:aes}), which serves one request at a time and computes each response over multiple cycles.

For each case study, we report on two aspects:
\begin{enumerate*}
  \item \emph{productivity}, where sequential semantics and flexible hazard resolution reduce development effort, and
  \item \emph{PPA}, where our designs outperform sequential-semantics HLS tools (PDL, Vitis HLS, Dynamatic) and match or exceed hand-written RTL and Kanagawa, both built under concurrent semantics.
\end{enumerate*}

\begin{table*}[t]
    \centering
    \scriptsize
\resizebox{\textwidth}{!}{%
    \begin{threeparttable}
    \caption{RISC-V CPU evaluation. The normalization baselines are highlighted in bold.}
    \label{cpu-eval}

    \begin{tabular}{@{}c|cccc|c|ccc|c|c@{}}
\toprule
               & \textbf{ISA} & \shortstack{\textbf{Regfile}\\\textbf{resolution}} & \shortstack{\textbf{PC}\\\textbf{resolution}} & \shortstack{\textbf{Pipeline}\tnote{\dag}\\\textbf{structure}} & \textbf{LOC} & \shortstack{\textbf{Max Freq.}\\\textbf{(MHz)}} & \shortstack{\textbf{CPI}\\\textbf{(Geomean)}} & \shortstack{\textbf{Norm.}\\\textbf{Runtime}} & \shortstack{\textbf{Area}\\\textbf{($\mu\mathrm{m}^2$)}} & \shortstack{\textbf{Power}\\\textbf{(mW)}} \\
\midrule
Ours Sim & \multicolumn{4}{c|}{-} & 1093 & \multicolumn{3}{c|}{-} & - & - \\
\midrule
\textbf{PDL}   & \multirow{3}{*}{RV32I}  & \multirow{3}{*}{Stall}     & \multirow{3}{*}{\shortstack{static\\prediction}} & \multirow{3}{*}{F/D/E/M/W}      & 250  & 1111 & 2.380 & 1.00 & 1896 (1.00x) & 4.33  (1.00x) \\
Sodor          &                         &                            &                                    &                                 & 1773 & 1250 & 2.190 & 0.82 & 969  (0.51x) & 2.58  (0.60x) \\
Ours           &                         &                            &                                    &                                 & 1092 & 1250 & 2.190 & 0.82 & 993  (0.52x) & 2.51  (0.58x) \\
\midrule
\textbf{PDL}   & \multirow{3}{*}{RV32I}  & \multirow{3}{*}{Bypass}    & \multirow{3}{*}{\shortstack{static\\prediction}} & \multirow{3}{*}{F/D/E/M/W}      & 262  & 833  & 1.395 & 1.00 & 1754 (1.00x) & 2.90  (1.00x) \\
Sodor          &                         &                            &                                    &                                 & 1778 & 1250 & 1.392 & 0.66 & 996  (0.57x) & 2.51  (0.86x) \\
Ours           &                         &                            &                                    &                                 & 1103 & 1250 & 1.392 & 0.66 & 1038 (0.59x) & 2.53  (0.87x) \\
\midrule
\textbf{PDL}   & \multirow{3}{*}{RV32I}  & \multirow{3}{*}{Bypass}    & \multirow{3}{*}{\shortstack{BHT\\(16 entries)}} & \multirow{3}{*}{F/D/E/M/W}      & 273  & 714  & 1.287 & 1.00 & 1964 (1.00x) & 2.74  (1.00x) \\
Sodor          &                         &                            &                                    &                                 & 1839 & 1250 & 1.280 & 0.57 & 1035 (0.53x) & 2.54  (0.93x) \\
Ours           &                         &                            &                                    &                                 & 1253 & 1250 & 1.280 & 0.57 & 1086 (0.55x) & 2.61  (0.95x) \\
\midrule
\textbf{PDL}   & \multirow{3}{*}{RV32I}  & \multirow{3}{*}{RenameRF}  & \multirow{3}{*}{\shortstack{static\\prediction}} & \multirow{3}{*}{F/D/E/M/W}      & 272  & 1111 & 1.617 & 1.00 & 3335 (1.00x) & 7.45  (1.00x) \\
Sodor          &                         &                            &                                    &                                 & 1836 & 1250 & 1.392 & 0.77 & 1771 (0.53x) & 4.47  (0.60x) \\
Ours           &                         &                            &                                    &                                 & 1164 & 1250 & 1.392 & 0.77 & 1833 (0.55x) & 4.64  (0.62x) \\
\midrule
\textbf{PDL}   & \multirow{3}{*}{RV32IM} & \multirow{3}{*}{Bypass}    & \multirow{3}{*}{\shortstack{static\\prediction}} & \multirow{3}{*}{F/D/E/M/W}      & 319  & 667  & 1.334 & 1.00 & 4724 (1.00x) & 6.45  (1.00x) \\
Sodor          &                         &                            &                                    &                                 & 1985 & 1250 & 2.086 & 0.83 & 1148 (0.24x) & 2.70  (0.42x) \\
Ours           &                         &                            &                                    &                                 & 1238 & 1250 & 2.086 & 0.83 & 1216 (0.26x) & 2.74  (0.42x) \\
\midrule
\textbf{Kanagawa} & \multirow{2}{*}{RV32I}  & \multirow{2}{*}{Bypass}    & \multirow{2}{*}{\shortstack{static\\prediction}} & \multirow{2}{*}{\shortstack{F1/F2D1/\\D2/EM/W}} & 1802 & 1250 & 1.546 & 1.00 & 1667 (1.00x) & 4.44  (1.00x) \\
Ours           &                         &                            &                                    &                                 & 1116 & 1250 & 1.546 & 1.00 & 1002 (0.60x) & 2.38  (0.54x) \\
\midrule
\textbf{Kanagawa} & \multirow{2}{*}{RV32I}  & \multirow{2}{*}{Bypass}    & \multirow{2}{*}{\shortstack{BTB\\(64 entries)}} & \multirow{2}{*}{\shortstack{F1/F2D1/\\D2/EM/W}} & 1893 & 1250 & 1.140 & 1.00 & 2751 (1.00x) & 7.93  (1.00x) \\
Ours           &                         &                            &                                    &                                 & 1253 & 1250 & 1.130 & 0.99 & 2244 (0.82x) & 6.13  (0.77x) \\
\bottomrule
\end{tabular}
    \begin{tablenotes}
        \footnotesize
        \item[*] PDL's LOC are shorter than ours; however, this is due to heavy inlining, no type definitions, and no Zicsr support.
        \item[**] Per-benchmark CPI and same-frequency synthesis results appear in \cref{appendix:riscv-eval}
        \item[\dag] F, D, E, M, and W denote the fetch, decode, execute, memory, and writeback stages.
        F1 selects the PC, and F2 fetches the instruction.
        D1 decodes it, and D2 reads the register file.
    \end{tablenotes}
    \end{threeparttable}
}
\end{table*}

\begin{table*}[t]
\centering
\scriptsize
\caption{A comparison of Countif histogram modules.}
\label{tab:evaluation-countif}
\resizebox{\textwidth}{!}{%
\begin{tabular}{c|r|r|rr|rr|rr}
\toprule
\multicolumn{1}{l|}{\multirow{2}{*}{}} & \multicolumn{1}{c|}{\multirow{2}{*}{Vitis}} & \multicolumn{1}{c|}{\multirow{2}{*}{Dynamatic}} & \multicolumn{2}{c|}{Static Sched}    & \multicolumn{2}{c|}{Dynamic Sched}   & \multicolumn{2}{c}{Speculative Exec} \\
\multicolumn{1}{l|}{}                  &                                             &                                                 & Kanagawa & \multicolumn{1}{c|}{Ours} & Kanagawa & \multicolumn{1}{c|}{Ours} & Kanagawa & \multicolumn{1}{c}{Ours}  \\
\midrule
LUT             & 361  & 9147 (25.34x) & 534 (1.48x)  & 493 (1.37x)  & 682 (1.89x)  & 506 (1.40x)  & 893 (2.47x)  & 754 (2.09x)  \\
FF              & 314  & 2345 (7.47x)  & 593 (1.89x)  & 297 (0.95x)  & 962 (3.06x)  & 419 (1.33x)  & 1398 (4.45x) & 794 (2.53x)  \\
BRAM            & 1    & 3 (3.00x)     & 1 (1.00x)    & 1 (1.00x)    & 1 (1.00x)    & 1 (1.00x)    & 1.5 (1.50x)  & 1 (1.00x)    \\
DSP             & 2    & 2 (1.00x)     & 2 (1.00x)    & 2 (1.00x)    & 2 (1.00x)    & 2 (1.00x)    & 2 (1.00x)    & 2 (1.00x)    \\
\midrule
Best-case Lat.  & 4135 & 1603 (0.39x)  & 4106 (0.99x) & 3073 (0.74x) & 529 (0.13x)  & 516 (0.12x)  & 1267 (0.31x) & 516 (0.12x)  \\
Worst-case Lat. & 4135 & 5706 (1.38x)  & 4106 (0.99x) & 3073 (0.74x) & 3595 (0.87x) & 2564 (0.62x) & 6895 (1.67x) & 2564 (0.62x) \\
\midrule
Lines of Code      & 24   & 28 (1.17x)    & 37 (1.54x)   & 22 (0.92x)   & 46 (1.92x)   & 23 (0.96x)   & 104 (4.33x)  & 39 (1.63x)   \\
\bottomrule
\end{tabular}%
}

\end{table*}

\subsection{RISC-V CPUs}
\label{evaluation:riscv}

For head-to-head comparison with PDL and Kanagawa, we implement a family of 5-stage RISC-V CPUs.
The CPUs vary the hazard resolution strategy (stalling, bypassing, dynamic branch prediction, and register renaming), the ISA (RV32I and RV32IM, both with Zicsr), and the stage division.
The baselines are PDL's public artifact~\cite{pdl}\footnote{We exclude XPDL~\cite{xpdl}, PDL's successor, because its public artifact only type-checks the CPU implementations and does not run benchmarks.}, Sodor~\cite{sodor}, and two Kanagawa RV32I cores~\cite{kanagawa}.\footnote{We exclude Kanagawa's RV32IM core, which supports multiplication but not division.}
As hand-written Chisel RTL of PDL's microarchitectures, Sodor serves as a performance upper bound.
One Kanagawa core uses static branch prediction, and the other uses a 64-entry branch target buffer (BTB).
Both cores split PDL's fetch and decode stages in two (F1/F2 and D1/D2 in \cref{cpu-eval}) and merge F2 with D1 and execute with memory.
Our two Kanagawa-style cores adopt the same stage division.
We synthesize all designs with Synopsys Design Compiler 2023.12~\cite{design_compiler} and the ASAP7 7nm PDK~\cite{asap7}.
We measure cycles per instruction (CPI) by simulating the generated RTL with Verilator~\cite{verilator} on CoreMark~\cite{eembc} and the MachSuite~\cite{machsuite} integer kernels.

\parhead{Productivity}
Productivity manifests in two forms:
\begin{enumerate*}
    \item \emph{stepwise development}, and
    \item \emph{debugging support}.
\end{enumerate*}
First, each feature of the CPU family takes a small code change to a 1,093-LOC Rust program that simulates an RV32I+Zicsr CPU:
\begin{enumerate*}
    \item 5-stage pipelining with static ``branch not taken'' prediction and data hazard stalls ($\pm$30 LOC),
    \item data bypassing ($\pm$33 LOC),
    \item dynamic branch prediction with a 16-entry BHT and 2-bit saturating counters ($\pm$155 LOC, of which 113 implement the BHT),
    \item register renaming ($\pm$69 LOC, of which 55 implement the renamed register file \code{RenameRF}), and
    \item a multiply-divide unit ($\pm$158 LOC).
\end{enumerate*}
The Kanagawa-style cores take two more code changes to the same simulator:
\begin{enumerate*}
    \item pipelining with their stage division ($\pm$48 LOC), and
    \item a 64-entry BTB ($\pm$86 LOC, of which 75 implement the BTB).
\end{enumerate*}
Second, we debug functionality at each step by executing the source as a Rust program, never the compiled pipeline.
Every change to an RTL design such as Sodor requires reasoning about concurrent module connections, which leave operation order implicit.
For performance debugging, we additionally provide a token-trace visualizer~\cite{supp}.
It shows each token's progress through the compiled pipeline cycle by cycle, helping designers locate stalls and adjust the pipelining strategy.

\parhead{PPA}
Our designs outperform PDL across all configurations, match Sodor, and use less area and power than Kanagawa at equal runtime (\cref{cpu-eval}).
Against PDL, they achieve $1.2\text{--}1.8\times$ lower runtime and $1.7\text{--}3.9\times$ smaller area.
Even with register renaming, which substantially increases design complexity, the core remains $1.8\times$ smaller and $1.3\times$ faster than PDL's.
Against Sodor at the same 1250~MHz clock frequency, our designs stay within $+3$ to $+6\%$ area and $-3$ to $+4\%$ power.
The parity stems from our unified compilation, which enables cross-module optimization across the main pipeline and \code{Var}s (\cref{rtl-compilation}).
Against Kanagawa, which also reaches 1250~MHz, our designs match its runtime within 1\% while occupying $1.2\text{--}1.7\times$ smaller area and drawing $1.3\text{--}1.9\times$ lower power.
The savings arise from how the two compilers implement a memory with multiple reads.
Kanagawa gives every read expression of a memory its own replica, and a write updates them all~\cite{kanagawa-guide}.
Each of its cores therefore holds two register-file replicas, one per source-register read.
Our compiler instead keeps a single base memory for \code{rf} and serves both reads from it through load circuits specialized to the pipeline structure (\cref{rtl-compilation:cpu}).

\subsection{Countif Histograms}
\label{eval:histogram}

For head-to-head comparison with Kanagawa, we implement countif histograms, its running example, under static, dynamic, and speculative hazard resolution strategies.
All three operate at address granularity over control- and data-dependent memory accesses.
The baselines, each reproduced from its public artifact, are Kanagawa~\cite{kanagawa}, Vitis HLS~\cite{vitis}, and Dynamatic~\cite{dynamatic}.
All designs target an Alveo U250 FPGA at a 200~MHz clock and are synthesized with AMD Vivado~2025.1.1.

\parhead{Productivity}
We switch among the three hazard resolution strategies through small code changes to the sequential source program, a flexibility neither Vitis HLS nor Dynamatic offers.

\parhead{PPA}
Our designs match or exceed Kanagawa's PPA under all three strategies and outperform Vitis HLS and Dynamatic (\cref{tab:evaluation-countif}).
The advantage over Kanagawa stems from a language-design tradeoff between precision and automation.
Our tool exposes stage boundaries directly, whereas Kanagawa exposes only a per-region concurrent-token count and delegates boundary placement to its scheduler.
Direct boundary control removes the latency and resource slack that a conservative scheduler retains.

\begin{table}[t]
\centering
\scriptsize
\caption{A comparison of AES cipher accelerators.}
\label{tab:evaluation-aes}
\begin{tabular}{c|ccccc}
\toprule
    & LUT & FF & \shortstack{Clock Freq.\\(MHz)} & \shortstack{Norm.\\Runtime} & \shortstack{Lines of\\Code} \\
\midrule
RTL                       & 3218 ($\times1.00$) & 2989 ($\times1.00$) & 128 ($\times1.00$) & 1.00 & 1844 ($\times1.00$)                 \\
\midrule
\multirow{2}{*}{Ours}     & 3175 ($\times0.99$) & 3002 ($\times1.00$) & 128 ($\times1.00$) & 0.99 & \multirow{2}{*}{549 ($\times0.30$)} \\
                            & 3246 ($\times1.01$) & 3002 ($\times1.00$) & 196 ($\times1.53$) & 0.65 &                                     \\
\midrule
\multirow{2}{*}{Ours Opt} & 2272 ($\times0.71$) & 673 ($\times0.23$)  & 128 ($\times1.00$) & 1.00 & \multirow{2}{*}{528 ($\times0.29$)} \\
                            & 2390 ($\times0.74$) & 673 ($\times0.23$)  & 179 ($\times1.40$) & 0.72 &                                     \\
\bottomrule
\end{tabular}

\end{table}

\subsection{AES Cipher Accelerator}
\label{eval:aes}

For applicability, we implement an AES cipher accelerator supporting both 128- and 256-bit keys.
The baseline is the SecWorks AES core~\cite{aes}, a widely used 1,844-LOC open-source Verilog implementation.%
\footnote{Kanagawa's AES-GCM design~\cite[\S6.2.2]{kanagawa} differs from our accelerator by adding the GCM authenticated-encryption mode and a pipelined dataflow.}
Both designs target an Artix~7 FPGA and are synthesized with AMD Vivado~2025.1.1.

\parhead{Productivity}
Productivity manifests in two forms: less code and surfaced optimizations.
First, our implementation takes $0.30\times$ the LOC of the reference and keeps the sequential algorithmic flow from key expansion to encryption and decryption, which the reference leaves implicit in concurrent module connections.
Second, the sequential code surfaces two optimizations:
\begin{enumerate*}
    \item \emph{path deduplication} shares state between the mutually exclusive encryption and decryption paths (an \code{if}/\code{else} branch in the source), and
    \item \emph{initialization elimination} removes redundant initialization logic for key memory, which the program always fills before use (via an element-wise store loop).
\end{enumerate*}

\parhead{PPA}
Compiled to the 128~MHz target of the reference, a direct port without either optimization matches its LUT and FF counts (\cref{tab:evaluation-aes}).
The port reaches a maximum frequency of 196~MHz, $1.53\times$ the reference's.
Applying both optimizations reduces LUT usage by $1.35\times$ and FF usage by $4.4\times$ at a maximum frequency of 179~MHz.

}
{\section{Related and Future Work}
\label{sec:related}

We discuss HLS abstractions for task- and data-level parallelism and hardware description languages (HDLs).
HLS abstractions for instruction-level parallelism, the focus of this work, are compared in \cref{sec:intro} and \cref{background:pdl-limitations}.

\parhead{HLS for task- and data-level parallelism}
Prior HLS work provides abstractions for task-level parallelism, the ability to execute multiple independent tasks simultaneously, and data-level parallelism, the ability to perform the same operation on multiple data elements in parallel.
We plan to extend our limited support for task- and data-level parallelism by integrating these abstractions.

For task-level parallelism, designers express concurrent computations using dataflow directives~\cite{vitis}, communicating processes~\cite{xls}, asynchronous functions~\cite{kanagawa}, or constructs for composing streaming tasks~\cite{spatial,allo}.
Compilers group computations into tasks to balance throughput and resource usage~\cite{scalehls,hida} and determine communication buffer depths to reduce stalls~\cite{vitis,vitis-fifo-sizing,spatial,allo}.

For data-level parallelism, designers use loop and array annotations~\cite{vitis,dahlia}, parallel loops and memory templates~\cite{spatial}, or composable program transformations~\cite{allo}.
Compilers select loop unrolling factors and memory banking configurations to improve performance within resource limits~\cite{spatial,scalehls,hida}.
Automatic scheduling explores hardware replication and reuse across cycles to reduce area while meeting a target throughput~\cite{aetherling}.

\parhead{HDLs}
Traditional HDLs~\cite{verilog,vhdl} expose wires, registers, and I/O ports for describing circuit structure.
With these low-level abstractions, designers implement instruction-, task-, and data-level parallelism by constructing pipelines, concurrently executing submodules, and replicated datapaths and banked memories.
However, coordinating data transfers and state updates in highly parallel designs is difficult because many interdependent operations execute concurrently.

HDL research addresses concurrency at the RTL level.
HDLs embedded in general-purpose languages~\cite{chisel,hardcaml,pyrtl1} use host-language metaprogramming to build reusable RTL generators for parameterized components and their connections.
Interface combinators~\cite{shakeflow,hazardflow} enable modular hardware design by composing concurrent components through latency-insensitive interfaces.
Guarded atomic actions~\cite{bluespec,cement2} express state updates as rules with atomic effects, and compilers generate control logic enforcing guards and scheduling constraints.
Type systems in Filament~\cite{filament} and Anvil~\cite{anvil} prevent components from using signals outside their valid time intervals.
Filament targets statically scheduled pipelines, whereas Anvil also supports interactions whose latency varies at runtime.
Our HLS tool instead takes a higher-level sequential program specifying an algorithm and its pipelining strategy and automatically generates RTL.

\parhead{Future work}
We are extending our work to develop application-class CPUs~\cite{cva6,boom}, AI accelerators~\cite{gemmini-dac}, and high-speed network cards~\cite{corundum,coyote}.
These designs require concurrency across modules, which our current tool cannot express because it accepts only sequential programs.
We are developing a message-passing extension in which sequentially specified modules communicate through channels, as in latency-insensitive design~\cite{li-interface}.
The extension would combine concurrent execution across modules with local visibility control for pipelining within each module.
For example, designers could implement out-of-order execution in application-class CPUs by specifying the scheduler and functional units as separate communicating sequential programs.
The scheduler would dispatch requests to the units out of instruction order.
}

{\section{Data-Availability Statement}
\label{appendix:sec:ae}

The artifact in the supplementary material~\cite{supp} contains the source code of our compiler, the designs of the three case studies (RISC-V CPUs, countif histograms, and an AES cipher accelerator), a token-trace visualizer, and evaluation scripts.
The scripts reproduce every table in the paper except for the RISC-V CPU synthesis results, which require Synopsys front-end tools (\code{lc\_shell} and \code{dc\_shell}).
We will submit all these materials for the Artifact Evaluation.
}

\bibliographystyle{ACM-Reference-Format}
\bibliography{reference}

\appendix
\clearpage
{

\section{PDL's Sequential Semantics}
\label{appendix:pdl_comparison}

This section provides detailed evidence for the fragility of PDL's sequential semantics guarantee.
We first enumerate the 18 rules that PDL and XPDL together impose on the programmer to preserve refinement between the pipeline and its sequential specification~(\cref{appendix:pdl:sequential_semantics}), then demonstrate that this rule set is unsound~(\cref{unsoundness}).

\subsection{Rules}
\label{appendix:pdl:sequential_semantics}

The sequential semantics guarantee of PDL and XPDL rests on 18 rules: 8 per-subsystem rules and 10 cross-subsystem interaction rules, of which the 8 rules governing the \code{except} and \code{commit} blocks arrive with XPDL's exceptions.
We enumerate these rules below, then show that they are unsound.
The rules keep PDL's own terminology, in which an in-flight instruction (\cref{background:token-model}) is called a thread.

\parhead{Per-subsystem rules}
PDL defines the following rules for using each subsystem independently:

\begin{enumerate}
  \item On every program path, each thread either spawns a single child thread (via a recursive \code{call} or \code{verify} statement) or terminates by \code{output}ting a value.
  \item On every program path, every speculative call is eventually resolved via a \code{verify} statement.
  \item On every program path, hazard locks must be used in the correct state-transition sequence of \code{reserve}, \code{block}, read/write, and \code{release}.
  \item All \code{reserve} operations must occur in ``in-order stages'', which are the stages where threads are guaranteed to pass in their program order.
  \item All \code{release} operations for write locks must occur in ``in-order stages''.
  \item All \code{reserve} operations for a hazard lock must execute atomically. That is, succeeding instructions' \code{reserve} operations should stall until the preceding instruction's \code{reserve} operation completes. PDL provides \code{lock region} construct to group multiple \code{reserve} operations into a critical section that executes atomically.
  \item A \code{verify} can only be called after \code{spec\_barrier()}.
  \item A \code{speccall} requires prior execution of either \code{spec\_check()} in the current pipeline stage or a \code{spec\_barrier()}.
\end{enumerate}

\parhead{Cross-subsystem interaction rules}
Because hazard locks, speculation, and exceptions are independent subsystems, their joint use requires 10 additional rules:

\begin{enumerate}
  \item A \code{reserve} requires the prior execution of either \code{spec\_check()} in the current pipeline stage or a \code{spec\_barrier()}.
  \item \code{release} on write locks can be called after \code{spec\_barrier()}.
  \item An instruction must \code{release} the lock before exiting the \code{except} block.
  \item Reads are forbidden from asynchronous memory or other pipelines at the end of the \code{except} block.
  \item A recursive \code{call} statement can only be in the last stage of the \code{except} block.
  \item \code{spec\_check}, \code{spec\_barrier}, and \code{spec\_call} statements are not allowed within the \code{except} or \code{commit} blocks.
  \item Users must call \code{spec\_barrier} before \code{except} or \code{commit} blocks.
  \item Users can not call \code{speccall} within the \code{except} or \code{commit} blocks.
  \item Write locks acquired in the pipeline body must be \code{release}d within the \code{commit} block and not before.
  \item No stateful operations are allowed in the \code{commit} block, with the exception of calling \code{release} to a lock. Stateful operations include spawning new instructions, acquiring locks, and speculation-related operations.
\end{enumerate}

\subsection{Unsoundness}
\label{unsoundness}

PDL's refinement guarantee between the pipeline and its sequential specification is unsound, in that a rule-compliant program accepted by the compiler can still produce a pipeline with incorrect hazard resolution.

\cref{fig:pdl-waw-result} illustrates a PDL program exhibiting a WAW hazard that originates from two conditional \code{release} operations.
Specifically, a thread with $\code{pc} = 1$ executes its \code{release} at L29, while the preceding thread with $\code{pc} = 0$ executes its \code{release} at L37.
The multiple pipeline stages separating these two \code{release} operations cause the thread with $\code{pc} = 0$
to release the lock and commit its \code{pc} after the thread with $\code{pc} = 1$, violating program order.

\begin{figure}[t]
  \centering
  \scriptsize
  \begin{subfigure}[c]{\columnwidth}
    \centering
\begin{minted}{rust}
pipe cpu(pc: uint<32>)[
    rf: uint<32>[5]<c2,s>(BypassQueue),
    imem: int<32>[32]<a,a>
]: bool {
    int<32> insn <- imem[pc];
    ---
    if (pc != u10<32>) {
        call cpu(pc + u1<32>);
    }
    ---
    uint<5> rd = u0<5>;
    ---
    uint<32> value = rf<a>[rd];
    if (pc == u10<32>) {
        print("Result is: %x", value);
        output(true);
    }
    ---
    if ((pc == 0) || (pc == 1)) {
        reserve(rf[rd], W);
        ---
        block(rf[rd]);
        ---
        rf[rd] <- pc;
    }
    ---
    if (pc == 1) {
        print("Release Start: %x", pc);
        release(rf[rd]);
        print("Release End: %x", pc);
    }
    ---
    ... // Pipeline stages
    ---
    if (pc == 0) {
        print("Release Start: %x", pc);
        release(rf[rd]);
        print("Release End: %x", pc);
    }
}
\end{minted}
    \caption{Rule-compliant PDL code with a WAW hazard.}
    \label{fig:pdl-waw-result}
  \end{subfigure}
  
  \begin{subfigure}[c]{\columnwidth}
    \centering
\begin{minted}{text}
Release Start: 00000001
Release End: 00000001
Release Start: 00000000
Release End: 00000000
Result is: 00000000
TIME                  435
\end{minted}
    \caption{Execution trace of \cref{fig:pdl-waw-result}}
    \label{fig:pdl-waw-trace}
  \end{subfigure}

  \caption{
    PDL code with a WAW hazard and its execution trace, with the unsound rule set fails to reject.
  }
  \label{fig:pdl-incomplete-rules}
\end{figure}

Because the PDL rules only verify whether each \code{release} statement is placed within in-order stages, this problematic code is not rejected.
The execution trace in \cref{fig:pdl-waw-trace} confirms the violation.
The ``Release Start'' for \code{pc = 1} (line 1) executes before the ``Release Start'' for \code{pc = 0} (line 3).
Consequently, ``Result is: 0000000'' shows that the final result is 0, whereas the result from the sequential specification should be 1.

\section{Extended RISC-V Evaluation}
\label{appendix:riscv-eval}

This section supplements \cref{evaluation:riscv} with per-benchmark CPI breakdowns (\cref{tab:riscv-cpi}) and same-frequency synthesis comparisons (\cref{tab:combined-riscv-performance}), followed by an analysis of the sources of PDL's overhead.

\begin{table*}[t]
    \centering
    \scriptsize
    \caption{Extended RISC-V CPU evaluation.}
    \label{tab:riscv-evaluation}

    \begin{subtable}[t]{\linewidth}
        \caption{CPI for benchmark programs (lower is better).}
        \label{tab:riscv-cpi}
        \centering
\resizebox{\textwidth}{!}{%

        \begin{tabular}{@{}c|ccc|ccccccccc|c@{}}
          \toprule
           & \textbf{ISA} & \shortstack{\textbf{Regfile}\\\textbf{resolution}} & \shortstack{\textbf{PC}\\\textbf{resolution}} & \textbf{coremark} & \textbf{aes} & \textbf{gemm} & \textbf{gemm-block} & \textbf{ellpack} & \textbf{kmp} & \textbf{nw} & \textbf{queue} & \textbf{radix} & \textbf{GeoMean} \\
          \midrule
          PDL            & \multirow{3}{*}{RV32I}  & \multirow{3}{*}{Stall}     & \multirow{3}{*}{static prediction} & 2.456 & 1.903 & 2.579 & 2.584 & 2.430 & 2.264 & 1.877 & 2.569 & 2.965 & 2.380 \\
          Sodor          &                         &                            &                                    & 2.177 & 1.851 & 2.386 & 2.402 & 2.177 & 2.018 & 1.672 & 2.334 & 2.932 & 2.190 \\
          Ours           &                         &                            &                                    & 2.177 & 1.850 & 2.386 & 2.402 & 2.177 & 2.018 & 1.672 & 2.334 & 2.932 & 2.190 \\
          \midrule
          PDL            & \multirow{3}{*}{RV32I}  & \multirow{3}{*}{Bypass}    & \multirow{3}{*}{static prediction} & 1.436 & 1.230 & 1.529 & 1.525 & 1.380 & 1.496 & 1.376 & 1.332 & 1.282 & 1.395 \\
          Sodor          &                         &                            &                                    & 1.441 & 1.201 & 1.530 & 1.525 & 1.381 & 1.497 & 1.377 & 1.332 & 1.283 & 1.392 \\
          Ours           &                         &                            &                                    & 1.441 & 1.200 & 1.530 & 1.525 & 1.381 & 1.497 & 1.377 & 1.332 & 1.283 & 1.392 \\
          \midrule
          PDL            & \multirow{3}{*}{RV32I}  & \multirow{3}{*}{Bypass}    & \multirow{3}{*}{BHT (16 entries)}  & 1.367 & 1.154 & 1.413 & 1.414 & 1.269 & 1.255 & 1.306 & 1.231 & 1.202 & 1.287 \\
          Sodor          &                         &                            &                                    & 1.362 & 1.156 & 1.408 & 1.406 & 1.254 & 1.255 & 1.271 & 1.235 & 1.202 & 1.280 \\
          Ours           &                         &                            &                                    & 1.362 & 1.155 & 1.408 & 1.406 & 1.254 & 1.255 & 1.271 & 1.234 & 1.202 & 1.280 \\
          \midrule
          PDL            & \multirow{3}{*}{RV32I}  & \multirow{3}{*}{Rename RF} & \multirow{3}{*}{static prediction} & 1.611 & 1.418 & 1.701 & 1.705 & 1.548 & 1.630 & 1.460 & 1.644 & 1.883 & 1.617 \\
          Sodor          &                         &                            &                                    & 1.441 & 1.201 & 1.530 & 1.525 & 1.381 & 1.497 & 1.377 & 1.332 & 1.283 & 1.392 \\
          Ours           &                         &                            &                                    & 1.441 & 1.200 & 1.530 & 1.525 & 1.381 & 1.497 & 1.377 & 1.332 & 1.283 & 1.392 \\
          \midrule
          PDL            & \multirow{3}{*}{RV32IM} & \multirow{3}{*}{Bypass}    & \multirow{3}{*}{static prediction} & 1.384 & 1.230 & 1.421 & 1.226 & 1.280 & 1.496 & 1.376 & 1.332 & 1.282 & 1.334 \\
          Sodor          &                         &                            &                                    & 1.579 & 1.201 & 5.852 & 4.503 & 4.249 & 1.497 & 1.377 & 1.332 & 1.283 & 2.086 \\
          Ours           &                         &                            &                                    & 1.579 & 1.200 & 5.852 & 4.503 & 4.249 & 1.497 & 1.377 & 1.332 & 1.283 & 2.086 \\
          \midrule
          Kanagawa       & \multirow{2}{*}{RV32I}  & \multirow{2}{*}{Bypass}    & \multirow{2}{*}{static prediction} & 1.615 & 1.264 & 1.795 & 1.788 & 1.571 & 1.743 & 1.565 & 1.355 & 1.327 & 1.546 \\
          Ours           &                         &                            &                                    & 1.616 & 1.261 & 1.795 & 1.788 & 1.571 & 1.743 & 1.565 & 1.355 & 1.326 & 1.546 \\
          \midrule
          Kanagawa       & \multirow{2}{*}{RV32I}  & \multirow{2}{*}{Bypass}    & \multirow{2}{*}{BTB (64 entries)}  & 1.252 & 1.090 & 1.290 & 1.290 & 1.114 & 1.013 & 1.165 & 1.044 & 1.043 & 1.140 \\
          Ours           &                         &                            &                                    & 1.232 & 1.078 & 1.272 & 1.273 & 1.093 & 1.013 & 1.164 & 1.037 & 1.042 & 1.130 \\
          \bottomrule
          \end{tabular}
          }%
    \end{subtable}

    \vspace{1em}

    \begin{subtable}[t]{\linewidth}
      \caption{Synthesis results. We report synthesis result of same frequency with the baseline and maximum frequency}
      \label{tab:combined-riscv-performance}
      \centering
      \begin{tabular}{@{}c|ccc|c|c|cc@{}}
        \toprule
         & \textbf{ISA} & \shortstack{\textbf{Regfile}\\\textbf{resolution}} & \shortstack{\textbf{PC}\\\textbf{resolution}} & \textbf{Target} & \shortstack{\textbf{Clock Freq.}\\\textbf{(MHz)}} & \textbf{Area ($\mu\mathrm{m}^2$)} & \textbf{Power (mW)} \\
        \midrule
        PDL    & \multirow{5}{*}{RV32I} & \multirow{5}{*}{Stall} & \multirow{5}{*}{static prediction} & -   & 1111 (0.9ns) & 1896 ($\times 1.00$) & 4.33 ($\times 1.00$) \\
        \multirow{2}{*}{Sodor} & & & & PDL & 1111 (0.9ns) & 960 ($\times 0.51$) & 2.22 ($\times 0.51$) \\
               & & & & Max & 1250 (0.8ns) & 969 ($\times 0.51$) & 2.58 ($\times 0.60$) \\
        \multirow{2}{*}{Ours} & & & & PDL & 1111 (0.9ns) & 995 ($\times 0.53$) & 2.16 ($\times 0.50$) \\
               & & & & Max & 1250 (0.8ns) & 993 ($\times 0.52$) & 2.51 ($\times 0.58$) \\
        \midrule
        PDL    & \multirow{5}{*}{RV32I} & \multirow{5}{*}{Bypass} & \multirow{5}{*}{static prediction} & -   & 833 (1.2ns) & 1754 ($\times 1.00$) & 2.90 ($\times 1.00$) \\
        \multirow{2}{*}{Sodor} & & & & PDL & 833 (1.2ns) & 976 ($\times 0.56$) & 1.63 ($\times 0.56$) \\
               & & & & Max & 1250 (0.8ns) & 996 ($\times 0.57$) & 2.51 ($\times 0.86$) \\
        \multirow{2}{*}{Ours} & & & & PDL & 833 (1.2ns) & 1024 ($\times 0.58$) & 1.64 ($\times 0.56$) \\
               & & & & Max & 1250 (0.8ns) & 1038 ($\times 0.59$) & 2.53 ($\times 0.87$) \\
        \midrule
        PDL    & \multirow{5}{*}{RV32I} & \multirow{5}{*}{Bypass} & \multirow{5}{*}{BHT (16 entries)} & -   & 714 (1.4ns) & 1964 ($\times 1.00$) & 2.74 ($\times 1.00$) \\
        \multirow{2}{*}{Sodor} & & & & PDL & 714 (1.4ns) & 1006 ($\times 0.51$) & 1.39 ($\times 0.51$) \\
               & & & & Max & 1250 (0.8ns) & 1035 ($\times 0.53$) & 2.54 ($\times 0.93$) \\
        \multirow{2}{*}{Ours} & & & & PDL & 714 (1.4ns) & 1063 ($\times 0.54$) & 1.43 ($\times 0.52$) \\
               & & & & Max & 1250 (0.8ns) & 1086 ($\times 0.55$) & 2.61 ($\times 0.95$) \\
        \midrule
        PDL    & \multirow{5}{*}{RV32I} & \multirow{5}{*}{Rename RF} & \multirow{5}{*}{static prediction} & -   & 1111 (0.9ns) & 3335 ($\times 1.00$) & 7.45 ($\times 1.00$) \\
        \multirow{2}{*}{Sodor} & & & & PDL & 1111 (0.9ns) & 1762 ($\times 0.53$) & 3.95 ($\times 0.53$) \\
               & & & & Max & 1250 (0.8ns) & 1771 ($\times 0.53$) & 4.47 ($\times 0.60$) \\
        \multirow{2}{*}{Ours} & & & & PDL & 1111 (0.9ns) & 1855 ($\times 0.56$) & 4.08 ($\times 0.55$) \\
               & & & & Max & 1250 (0.8ns) & 1833 ($\times 0.55$) & 4.64 ($\times 0.62$) \\
        \midrule
        PDL    & \multirow{5}{*}{RV32IM} & \multirow{5}{*}{Bypass} & \multirow{5}{*}{static prediction} & -   & 667 (1.5ns) & 4724 ($\times 1.00$) & 6.45 ($\times 1.00$) \\
        \multirow{2}{*}{Sodor} & & & & PDL & 667 (1.5ns) & 1098 ($\times 0.23$) & 1.39 ($\times 0.21$) \\
               & & & & Max & 1250 (0.8ns) & 1148 ($\times 0.24$) & 2.70 ($\times 0.42$) \\
        \multirow{2}{*}{Ours} & & & & PDL & 667 (1.5ns) & 1168 ($\times 0.25$) & 1.40 ($\times 0.22$) \\
               & & & & Max & 1250 (0.8ns) & 1216 ($\times 0.26$) & 2.74 ($\times 0.42$) \\
        \midrule
        Kanagawa & \multirow{2}{*}{RV32I} & \multirow{2}{*}{Bypass} & \multirow{2}{*}{static prediction} & -   & 1250 (0.8ns) & 1667 ($\times 1.00$) & 4.44 ($\times 1.00$) \\
        Ours   & & & & -   & 1250 (0.8ns) & 1002 ($\times 0.60$) & 2.38 ($\times 0.54$) \\
        \midrule
        Kanagawa & \multirow{2}{*}{RV32I} & \multirow{2}{*}{Bypass} & \multirow{2}{*}{BTB (64 entries)} & -   & 1250 (0.8ns) & 2751 ($\times 1.00$) & 7.93 ($\times 1.00$) \\
        Ours   & & & & -   & 1250 (0.8ns) & 2244 ($\times 0.82$) & 6.13 ($\times 0.77$) \\
        \bottomrule
        \end{tabular}
  \end{subtable}
\end{table*}

\parhead{Sources of PDL overhead}
\cref{tab:combined-riscv-performance} provides a same-frequency comparison between PDL's 5-stage RISC-V CPU and ours under various configurations.
This comparison is unfavorable to our design, as ours additionally supports exceptions and the Zicsr ISA extension, which PDL's design lacks.
When targeting PDL's clock speeds, our processor uses 42--48\% less area and 43--50\% less power across the rv32i configurations.
The gap instantiates the two implementation overheads of \cref{background:pdl-limitations}.
First, PDL's subsystems are four redundant ordering systems, each adding per-instruction metadata as pipeline registers.
The speculation subsystem additionally maintains its speculation table because a speculative instruction keeps executing until it voluntarily checks its status at \code{spec\_check} or \code{spec\_barrier}~(\cref{background:token-model}).
Second, PDL performs unnecessary order comparison.
Each lock compiles to a hand-written RTL module unaware of the pipeline structure, so it tracks the program order of all in-flight instructions with runtime comparators for priority calculation and arbitration.
Our program-order analysis (\cref{rtl-compilation:cpu}) instead decides program order statically wherever possible.
In our RISC-V case studies, it decides every pair, so the compiler generates direct bypass wires and no comparators.

}

\end{document}